\documentstyle[psfig]{mn}

\newif\ifAMStwofonts

\newcommand{\simgt}{\lower.5ex\hbox{$\; \buildrel > \over \sim \;$}}
\newcommand{\simlt}{\lower.5ex\hbox{$\; \buildrel < \over \sim \;$}}

\newcommand{\mbkt}[1]{\left\{#1\right\}}
\newcommand{\sbkt}[1]{\left(#1\right)}

\newcommand{\msun}{\,{\rm M}_{\odot}}

\ifoldfss
  \ifCUPmtlplainloaded \else
    \NewTextAlphabet{textbfit} {cmbxti10} {}
    \NewTextAlphabet{textbfss} {cmssbx10} {}
    \NewMathAlphabet{mathbfit} {cmbxti10} {} 
    \NewMathAlphabet{mathbfss} {cmssbx10} {} 
  \fi
  \ifAMStwofonts
    \ifCUPmtlplainloaded \else
      \NewSymbolFont{upmath} {eurm10}
      \NewSymbolFont{AMSa} {msam10}
      \NewMathSymbol{\upi}     {0}{upmath}{19}
      \NewMathSymbol{\umu}     {0}{upmath}{16}
      \NewMathSymbol{\upartial}{0}{upmath}{40}
      \NewMathSymbol{\leqslant}{3}{AMSa}{36}
      \NewMathSymbol{\geqslant}{3}{AMSa}{3E}

      \let\leq=\leqslant 
      \let\geq=\geqslant 
    \fi
  \fi
\fi 

\ifnfssone
  \newmathalphabet{\mathit}
  \addtoversion{normal}{\mathit}{cmr}{m}{it}
  \addtoversion{bold}{\mathit}{cmr}{bx}{it}
  \newmathalphabet{\mathbfit} 
  \addtoversion{normal}{\mathbfit}{cmr}{bx}{it}
  \addtoversion{bold}{\mathbfit}{cmr}{bx}{it}
  \newmathalphabet{\mathbfss} 
  \addtoversion{normal}{\mathbfss}{cmss}{bx}{n}
  \addtoversion{bold}{\mathbfss}{cmss}{bx}{n}
  \ifAMStwofonts
    \ifCUPmtlplainloaded \else
      \UseAMStwoboldmath
      \makeatletter
      \new@mathgroup\upmath@group
      \define@mathgroup\mv@normal\upmath@group{eur}{m}{n}
      \define@mathgroup\mv@bold\upmath@group{eur}{b}{n}
      \edef\UPM{\hexnumber\upmath@group}
      \new@mathgroup\amsa@group
      \define@mathgroup\mv@normal\amsa@group{msa}{m}{n}
      \define@mathgroup\mv@bold\amsa@group{msa}{m}{n}
      \edef\AMSa{\hexnumber\amsa@group}
      \makeatother
      \mathchardef\upi="0\UPM19
      \mathchardef\umu="0\UPM16
      \mathchardef\upartial="0\UPM40
      \mathchardef\leqslant="3\AMSa36
      \mathchardef\geqslant="3\AMSa3E

      \let\leq=\leqslant 
      \let\geq=\geqslant 
    \fi
  \fi
\fi 

\ifnfsstwo
  \DeclareMathAlphabet{\mathbfit}{OT1}{cmr}{bx}{it}
  \SetMathAlphabet\mathbfit{bold}{OT1}{cmr}{bx}{it}
  \DeclareMathAlphabet{\mathbfss}{OT1}{cmss}{bx}{n}
  \SetMathAlphabet\mathbfss{bold}{OT1}{cmss}{bx}{n}
  \ifAMStwofonts
    \ifCUPmtlplainloaded \else
      \DeclareSymbolFont{UPM}{U}{eur}{m}{n}
      \SetSymbolFont{UPM}{bold}{U}{eur}{b}{n}
      \DeclareSymbolFont{AMSa}{U}{msa}{m}{n}
      \DeclareMathSymbol{\upi}{0}{UPM}{"19}
      \DeclareMathSymbol{\umu}{0}{UPM}{"16}
      \DeclareMathSymbol{\upartial}{0}{UPM}{"40}
      \DeclareMathSymbol{\leqslant}{3}{AMSa}{"36}
      \DeclareMathSymbol{\geqslant}{3}{AMSa}{"3E}

      \let\leq=\leqslant 
      \let\geq=\geqslant 
    \fi
  \fi
\fi 

\ifCUPmtlplainloaded \else
  \ifAMStwofonts \else 
    \def\upi{\pi}
    \def\umu{\mu}
    \def\upartial{\partial}
  \fi
\fi

\title[Criteria for the Formation of Population III Objects] {Criteria
for the Formation of Population III Objects in the Ultraviolet
Background Radiation} \author[T. Kitayama et
al.]{T. Kitayama,$^{1}$\footnotemark H. Susa,$^{2}$ M. Umemura$^{2}$
and S. Ikeuchi$^{3}$ \\ $^{1}$ Department of Physics, Tokyo
Metropolitan University, Hachioji, Tokyo 192-0397, Japan\\ $^{2}$
Center for Computational Physics, University of Tsukuba, Tsukuba
305-8577, Japan\\ $^{3}$ Department of Physics, Nagoya University,
Chikusa-ku, Nagoya 464-8602, Japan}

\date{Accepted for publication in MNRAS.}

\pagerange{\pageref{firstpage}--\pageref{lastpage}}
\pubyear{2001}

\begin{document}

\maketitle

\label{firstpage}

\begin{abstract}
We explore possibilities of collapse and star formation in Population
III objects exposed to the external ultraviolet background (UVB)
radiation.  Assuming spherical symmetry, we solve self-consistently
radiative transfer of photons, non-equilibrium H$_2$ chemistry, and
gas hydrodynamics.  Although the UVB does suppress the formation of
low mass objects, the negative feedback turns out to be weaker than
previously suggested. In particular, the cut-off scale of collapse
drops significantly below the virial temperature $T_{\rm vir}\sim
10^4$K at weak UV intensities ($J_{21}\simlt 10^{-2}$), due to both
self-shielding of the gas and H$_2$ cooling. Clouds above this cut-off
tend to contract highly dynamically, further promoting self-shielding
and H$_2$ formation. For plausible radiation intensities and spectra,
the collapsing gas can cool efficiently to temperatures well below
$10^4$K before rotationally supported and the final H$_2$ fraction
reaches $\sim 10^{-3}$.

Our results imply that star formation can take place in low mass
objects collapsing in the UVB. The threshold baryon mass for star
formation is $\sim 10^9 \msun$ for clouds collapsing at redshifts $z
\simlt 3$, but drops significantly at higher redshifts. In a
conventional cold dark matter universe, the latter coincides roughly
with that of the 1$\sigma$ density fluctuations. Objects near and
above this threshold can thus constitute `building blocks' of
luminous structures, and we discuss their links to dwarf
spheroidal/elliptical galaxies and faint blue objects. These results
suggest that the UVB can play a key role in regulating the star
formation history of the Universe.
\end{abstract}


\footnotetext{E-mail: tkita@phys.metro-u.ac.jp}

\section{Introduction}
\label{sec:intro}

The existence of an intense ultraviolet background (UVB) radiation,
inferred from observations of QSO absorption spectra (Gunn \& Peterson
1965; Bajtlik, Duncan \& Ostriker 1988), is likely to have prominent
impacts on galaxy formation. While its origin could be attributed to
QSOs or young galaxies appeared at high redshifts (e.g., Couchman
1985; Miralda-Escude \& Ostriker 1990; Sasaki \& Takahara 1994;
Fukugita \& Kawasaki 1994), the UVB once produced photoionizes the
intergalactic medium and alters the subsequent growth of cosmic
structures. These secondary structures emerged in the UVB are further
responsible for producing the radiation field, which can in turn
affect the development of next generation objects. It is therefore
crucial to determine quantitatively the consequences of radiative
feedback on the formation of early generation (hereafter Population
III) objects, in order to link the build-up of structures with the
thermal history of the universe.

It has long been suggested that the formation of subgalactic objects
is suppressed via photoionization and heating caused by the UVB
(Umemura \& Ikeuchi 1984; Ikeuchi 1986; Rees 1986; Bond, Szalay \&
Silk 1988; Efstathiou 1992; Babul \& Rees 1992; Murakami \& Ikeuchi
1993; Chiba \& Nath 1994; Zhang, Anninos \& Norman 1995; Thoul \&
Weinberg 1996, among others). Most of previous studies, however,
assumed for simplicity that the medium is optically thin against the
external radiation. Several attempts have been made recently to take
into account radiative transfer of photons (e.g., Kepner, Babul \&
Spergel 1997; Tajiri \& Umemura 1998; Abel, Norman \& Madau 1999;
Barkana \& Loeb 1999; Razoumov \& Scott 1999; Gnedin 2000b; Susa \&
Umemura 2000a; Nakamoto, Umemura \& Susa 2001; Ciardi et al. 2001). In
our previous papers (Kitayama \& Ikeuchi 2000, hereafter Paper I;
Kitayama et al. 2000, hereafter Paper II), we have studied the
evolution of spherical clouds exposed to the UVB, solving explicitly
radiative transfer of ionizing photons and hydrodynamics. We have
found that the collapse of objects with circular velocities below
$V_{\rm c} \sim 15 - 40$ km s$^{-1}$ can be prohibited completely even
if self-shielding of gas is taken into account. We have also noted
that a spherical cloud above this threshold undergoes run-away
collapse in the dark matter potential and does not settle into
hydrostatic equilibrium.  For plausible intensities and spectra of the
UVB, the central region of such a cloud is shown to get self-shielded
against photoionization before shrinking to the rotation barrier. A
question still remains as to whether star formation can take place
efficiently in the cloud collapsing under the UVB.

A key element in this context is molecular hydrogen (H$_2$). In order
for stars to form, the gas needs to radiate energy efficiently and
cool down to temperatures well below $T=10^4$K. In the gas of
primordial composition without metals, cooling processes at this
temperature are almost solely dominated by rotational-vibrational line
excitation of H$_2$ (e.g., Peebles \& Dicke 1968; Hirasawa 1969;
Matsuda, Sato \& Takeda 1969).  Formation and destruction of H$_2$,
however, are very sensitive to the presence of a radiation field. For
instance, H$_2$ molecules are easily dissociated by photons in the
so-called Lyman-Werner (LW, 11.2--13.6 eV) bands (Stecher \& Williams
1967; Haiman, Rees \& Loeb 1997; Ciardi et al 2000b) or via collisions
with ions. On the other hand, the gas-phase formation of H$_2$ is
promoted by an enhanced ionized fraction in the post-shock gas (Shapiro
\& Kang 1987; Kang \& Shapiro 1992) or in weakly photoionized media
(Haiman, Rees \& Loeb 1996b). In addition, self-shielding taking place
during dynamical collapse can also aid H$_2$ cooling (Susa \& Kitayama
2000).  It is thus by no means trivial in what circumstances the UVB
has positive or negative feedback on star formation in Population III
objects.

In this paper, by including H$_2$ chemistry consistently with
radiative transfer of UV photons, we extend our previous analyses
(Papers I and II) and study the fate of Population III objects exposed
to the external UVB. Assuming spherical symmetry, we solve the
radiative transfer equation, non-equilibrium chemical reactions
including H$_2$ formation and destruction, and dynamics of baryonic
gas and dark matter. Particular attention is paid to the possibilities
of both collapse and H$_2$ cooling in the presence of UV radiation
fields. We explore the evolution of objects with wide ranges of
masses, including those with the virial temperature $T_{\rm vir} <
10^4$K ($V_{\rm c} < 17$ km s$^{-1}$), which have been left out in our
previous analyses neglecting H$_2$ formation. A similar problem was
investigated earlier by Kepner et al. (1997) and Haiman, Abel \& Rees
(2000), assuming that collapsed clouds are in approximate hydrostatic
equilibrium. As mentioned above and shown explicitly in this paper,
the gas in collapsed objects is far from static and its physical state
changes highly dynamically. We also note that our analysis is
complementary to that of Susa \& Umemura (2000a,b), who explored the
pregalactic pancake collapse in plane-parallel geometry and studied
the fate of objects far above the Jeans scale. We pay particular
attention to the development of smaller objects near the Jeans scale,
which are likely to evolve spherically to a first-order approximation. 
A crucial distinction between the earlier works and ours is that the
gas density rises rapidly along with the spherical collapse, in
proportion to the cube of scale length, providing greater chances of
self-shielding and H$_2$ cooling.

This paper is organized as follows. Section 2 describes the numerical
model used in the paper. We present the simulation results in Section
3 and discuss their implications on galaxy formation in Section
4. Finally, Section 5 summarizes our conclusions. Throughout the
paper, we assume the density parameter $\Omega_0=0.3$, the
cosmological constant $\lambda_0=0.7$, the Hubble constant
$h_{100}=H_0$/(100 km s$^{-1}$ Mpc$^{-1})=0.7$, and the baryon density
parameter $\Omega_{\rm b} = 0.05$.

\section{Model}
\label{sec:model}

\subsection{Numerical Scheme}

We simulate the evolution of a spherically symmetric cloud exposed to
the external UVB.  The cloud is a mixture of baryonic gas and dark
matter with the mass ratio of $\Omega_{\rm b} : \Omega_0 - \Omega_{\rm
b} = 1 : 5$. The basic equations are summarized in Section 2.1. of
Paper I. One modification is that the acceleration term due to the
cosmological constant $ + \lambda_0 H_0^2 r^2$ is added to the right
hand side of momentum equations. They are solved with the
second-order-accurate Lagrangian finite-difference scheme described in
Bowers \& Wilson (1991) and Thoul \& Weinberg (1995). Shocks are
treated with an artificial tensor viscosity (Tscharnuter \& Winkler
1979). The number of mass shells is $N_{\rm b}=300$ for baryonic gas
and $N_{\rm d}=10,000$ for dark matter. Our code reproduces accurately
the similarity solutions for the adiabatic accretion of collisional
gas and the pressureless collapse around a point-mass perturbation
(Bertschinger 1985). We have also checked that our results are
essentially unchanged when the shell numbers are doubled or halved.

At each time-step of the simulation, we solve self-consistently the
radiative transfer equation, chemical reactions, and energy equation
in the following three steps. Repeating these procedures, the
radiation field and the internal state of the gas are determined
iteratively until the abundance of each species and the internal
energy in each mesh converge within an accuracy of 0.1\%.

First, the direction/frequency-dependent radiative transfer is worked
out for ionizing photons with energies $h \nu \geq 13.6$ eV, with the
method devised by Tajiri \& Umemura (1998), in which both absorption
and emission by hydrogen are explicitly taken into account. At each
radial point, angular integration of the transfer equation is done
over at least 20 bins in $\theta = 0 -\pi$, where $\theta$ is the
angle between the light ray and the radial direction. This is achieved
by handling 400--1,000 impact parameters for light rays.
Self-shielding of H$_2$ against the LW photons (11.2--13.6 eV) is
evaluated from the self-shielding function of Draine \& Beltoldi
(1996), considering the direction dependence of the H$_2$ column
density as in Kepner et al. (1997). The gas is assumed to be optically
thin against the photons with $h \nu < 11.2$eV, which only contribute
to destructions of H$^-$ and H$_2^+$.

Secondly, non-equilibrium chemical reactions are solved in each mesh
for the species e, H, H$^+$, H$^-$, H$_2$, H$_2^+$. Hydrogen molecules
form mainly via the creation of the intermediates H$^-$ and H$_2^+$
(eqs [1.1]--[1.4] of Shapiro \& Kang 1987). We neglect helium for the
following reasons: (i) it only changes the ionization degree of the
gas up to $\sim 10 \%$ (Osterbrock 1989; Susa \& Umemura 2000a); (ii)
helium line cooling is unimportant at the typical temperatures ($T
\simlt 10^4$K) of the objects presently studied; and (iii)
photoionization heating of helium does not alter our main results as
discussed in Section \ref{sec:signif}.  Unless stated explicitly,
reaction rates are taken from the recent compilation of Galli \& Palla
(1998). The coefficients for photoionization of H and
photodissociation of H$_2$ are computed from the results of the
radiative transfer mentioned above. Photodetachment of H$^-$ and
photodissociation of H$_2^+$ are also taken into account using the
cross sections given in Tegmark et al. (1997) and Stancil (1994) for
the photons with $h \nu \geq 0.74$ eV and $h \nu \geq 0.062$ eV,
respectively. These two processes, particularly the former, can
suppress H$_2$ formation by destroying its intermediates.

Finally, the energy equation is solved including UV heating and
radiative cooling due to collisional ionization, collisional
excitation, recombination, thermal bremsstrahlung, Compton scattering
with the cosmic microwave background (CMB) radiation, and
rotational-vibrational excitation of H$_2$. The UV heating rate is
obtained directly from the results of the radiative transfer. The
atomic cooling rates are taken from Fukugita \& Kawasaki (1994) and
molecular cooling rates from Galli \& Palla (1998).

For comparison with previous approaches, we also perform simpler
calculations in the following two ways. One assumes at each time-step
an optically thin medium and ionization equilibrium among e, H and
H$^+$, neglecting H$_2$, H$^-$ and H$_2^+$. The other solves radiative
transfer of ionizing photons, but neglects H$_2$, H$^-$ and H$_2^+$.
The latter corresponds to the analysis of Paper II. Hereafter we refer
to the former as the `optically thin calculation', the latter as the
`no H$_2$ calculation', and the one described in the previous
paragraphs as the `full calculation'.  Unless otherwise stated, the
full calculation is adopted.

\subsection{External UVB}

\begin{figure}
\begin{center}
   \leavevmode\psfig{figure=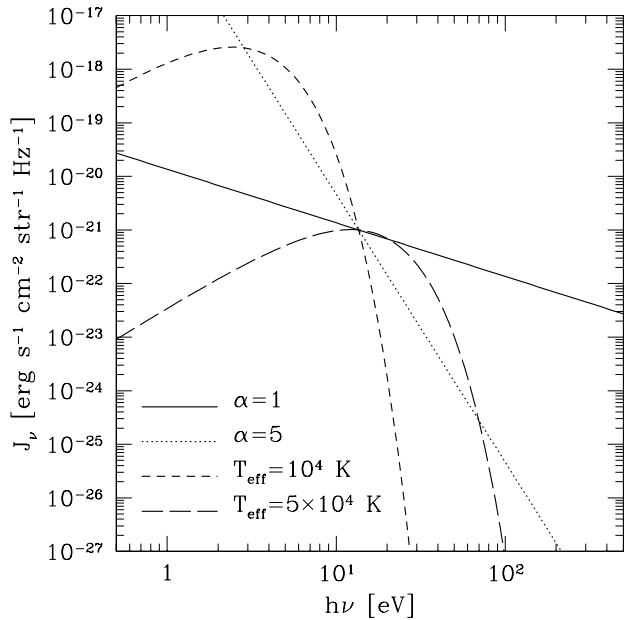,width=8.5cm}
\end{center}
\caption{Model spectra of the external UVB with $\alpha=1$ (solid),
$\alpha=5$ (dotted), $T_{\rm eff} = 10^4$K (short dashed), and $T_{\rm
eff} = 5 \times 10^4$K (long dashed), all normalized to give
$J_{21}=1$.}
\label{fig:bbspect}
\end{figure}

The external UVB is assumed to be isotropic and to have either a
power-law or black-body spectrum:
\begin{equation}
  J^{\rm ext}_\nu \propto \left\{\begin{array}{l}
   \nu^{-\alpha}, \\
   \frac{\nu^3}{\exp(h \nu /k_{\rm B}T_{\rm eff}) - 1 }, \\
    \end{array} \right.  
\label{eq:uvb}
\end{equation}
where $k_{\rm B}$ is the Boltzmann constant.  We adopt the spectral
index $\alpha=1$ and the effective temperature $T_{\rm eff}=10^4$K, to
mimic the spectra of QSOs and stars, respectively. The above spectrum
is normalized to give the intensity $J_{21}$, in units of
$10^{-21}$erg s$^{-1}$ cm$^{-2}$ str$^{-1}$ Hz$^{-1}$, at the Lyman
limit of hydrogen ($h \nu =13.6$eV).

Fig. 1 illustrates the model spectra described above with
$J_{21}=1$. Also shown for reference are those with $\alpha=5$ and
$T_{\rm eff}=5 \times 10^4$K. In the present analyses, the ratio
between the fluxes below and above the Lyman limit is a key parameter
to control the radiative feedback on the gas evolution (e.g., Ciardi,
Ferrara \& Abel 2000a). To help quantitative comparisons among
different spectral shapes, we list in Table 1 the
photon-number-weighted average intensities of these spectra:
\begin{equation}
\langle J \rangle = \frac{\int
J_{\nu}^{\rm ext}/h\nu ~d\nu} {\int 1/h\nu ~d\nu} \times \frac{1}
{\mbox{$10^{-21}$erg s$^{-1}$ cm$^{-2}$ str$^{-1}$ Hz$^{-1}$}},
\label{eq:jb}
\end{equation}
where the average is taken over the energies relevant for destruction
of both H$_2$ and H$^{-}$ (0.74--13.6 eV), dissociation of only H$_2$
(11.2--13.6 eV), and ionization of H (13.6--$10^4$ eV). Notice that
for a given value of $J_{21}$ the spectrum with $T_{\rm eff}=10^4$K
has considerably less ionizing ($> 13.6$eV) photons and more H$_2$ and
H$^{-}$ destroying photons ($< 13.6$eV) than the one with $\alpha=1$.

\begin{table}
\caption{Average intensities $\langle J \rangle$ defined by equation
(\protect\ref{eq:jb}\protect) in specific energy ranges. All spectra
are normalized to give $J_{21}=1$.}
\label{tab:uv}
\begin{center}
\begin{tabular}{lccc} \\
\hline \\[-6pt] spectrum 
& 0.74--13.6 eV & 11.2--13.6 eV & 13.6--$10^4$ eV 
   \\[4pt]\hline \\[-6pt]
$\alpha=1$& 6.0 & 1.1  & 0.15 \\  
$\alpha=5$& $1.4\times 10^5$ & 1.7 & $3.0 \times 10^{-2}$\\  
$T_{\rm eff}=10^4$ K & $1.3 \times 10^{3}$ &  3.8 & $1.1  \times 10^{-2}$\\
$T_{\rm eff}=5 \times 10^4$ K & 0.39 & 1.0 & $8.6 \times 10^{-2}$ \\[4pt] 
\hline 
\end{tabular} 
\end{center}
\end{table}

Observations of the proximity effect in the Ly$\alpha$ forest suggest
$J_{21} = 10^{\pm 0.5}$ at redshift $z=1.7-4.1$ (Bajtlik et al. 1988;
Bechtold 1994; Giallongo et al. 1996; Cooke, Espey \& Carswell 1997),
but its value is still highly uncertain at other redshifts. Recent CMB
power spectrum measurement by MAXIMA has been used to constrain the
redshift of reionization to be $z \simgt 8$ at the 2$\sigma$
confidence level (Schmalzing, Sommer-Larsen \& G\"{o}tz 2000). In this
paper, we fiducially fix the onset of the UVB at $z_{\rm UV} =20$ and
consider at $z \leq z_{\rm UV}$ constant $J_{21}$ of an arbitrary
value. To take into account uncertainties in the reionization history,
we also examine some cases with $z_{\rm UV}=10$ and $50$, and evolving
$J_{21}$ of the form:
\begin{equation}
  J_{21} = \left\{\begin{array}{ll}
       \exp[-(z-5)] &  5 \leq z \leq 20 \\
      1 &  3 \leq z \leq 5 \\ 
      \sbkt{\frac{1+z}{4}}^4 &  0 \leq z \leq 3, \\ 
     \end{array} \right.
\label{eq:uvevl}
\end{equation}
where the evolution at $z>5$ is roughly consistent with recent models
of reionization of the universe (Ciardi et al 2000b; Umemura, Nakamoto
\& Susa 2000).

\subsection{Initial and boundary conditions}
Initial and boundary conditions are similar to those in Sections
2.3. and 2.4. of Paper I. In the following, we give their brief
summary and state any modifications and additions.

We start the simulations when the overdensity of a cloud is still in
the linear regime. The initial overdensity profile for both baryon and
dark matter is $\delta_{\rm i}(r) = \delta_{\rm i}(0)\sin(kr)/kr$,
where $k$ is the comoving wave number, and the central overdensity
$\delta_{\rm i}(0)$ is fixed at 0.2.  Only the growing mode is
considered. The outer boundary is taken at the first minimum of
$\delta_{\rm i}(r)$, i.e., $kr = 4.4934$, within which the volume
averaged overdensity $\bar{\delta}_{\rm i}(<r)$ vanishes. The
characteristic mass of a cloud $M_{\rm cloud}$ is defined as the
baryon mass enclosed within the first zero of $\delta_{\rm i}(r)$,
i.e., $kr = \pi$, which is 0.363 of the total baryon mass (Haiman,
Thoul \& Loeb 1996a). We define the central collapse redshift $z_{\rm
c0}$ and the cloud collapse redshift $z_{\rm c}$, respectively, as
epochs at which the inner-most gas shell and the shell enclosing
$M_{\rm cloud}$ would collapse to the center in the absence of thermal
pressure. These two redshifts are related by $1+z_{\rm c0} \simeq 2.7
(1+z_{\rm c})$ (strictly speaking, this relation is correct in the
Einstein-de Sitter universe, but it still gives a reasonable
approximation in the present case at $z_{\rm c} \simgt 1$). We will
often characterize a cloud by its circular velocity $V_{\rm c}$ and
virial temperature $T_{\rm vir}$, defined in terms of $z_{\rm c}$ and
$M_{\rm cloud}$ as
\begin{eqnarray}
 V_{\rm c} &=& 15.9 \sbkt{\frac{M_{\rm cloud} \Omega_0/\Omega_{\rm b}}
{10^9 h_{100}^{-1} \msun}}^{1/3} \nonumber \\
&& \times \mbkt{\frac{\Delta_{\rm c}(z_{\rm c})}{18\pi^2}}^{1/6}
(1+z_{\rm c})^{1/2} ~\mbox{  km s$^{-1}$}, 
\label{eq:vc}
\end{eqnarray}
\begin{eqnarray}
T_{\rm vir} &=& 9.09 \times 10^3  \sbkt{\frac{\mu}{0.59}}
\sbkt{\frac{M_{\rm cloud} \Omega_0/\Omega_{\rm b}}
{10^9 h_{100}^{-1} \msun}}^{2/3} \nonumber \\
&&\times  \mbkt{\frac{\Delta_{\rm c}(z_{\rm c})}{18\pi^2}}^{1/3}(1+z_{\rm c}) ~\mbox{  K}, 
\label{eq:tvir}
\end{eqnarray}
where $\mu$ is the mean molecular weight in units of the proton mass
$m_{\rm p}$, and $\Delta_{\rm c}(z_{\rm c})$ is the mean overdensity of a
virialized object predicted from the spherical infall model (Peebles
1980; Kitayama \& Suto 1996). Using circular velocity or virial
temperature to denote the size of an object, we can minimize the
dependences of our simulation results on assumed cosmological
parameters.

The initial abundance of each species is taken to be its cosmological
post-recombination value (e.g., Galli \& Palla 1998), i.e., $X_{\rm
e}=10^{-4}$, $X_{{\rm H}^-} =10^{-12}$, $X_{\rm H2}=10^{-6}$, and
$X_{\rm{H2}^+}=10^{-13}$, where $X_{\rm i}\equiv n_{\rm i}/n_{\rm H}$
is the fraction of the species i with respect to the total number of
hydrogen atoms. Our results are insensitive to the precise values of
initial abundances. In the optically thin and no H$_2$ calculations,
$X_e=10^{-4}$ is assumed before the gas is exposed to the UVB or
experiences shocks.

In this paper, we mainly focus on the evolution of the central gas of
a cloud until it reaches the rotation radius specified by the
dimensionless spin parameter at turn-around $\lambda_{\rm ta}$:
\begin{equation}
r_{\rm rot}= 0.03 \sbkt{\frac{\Omega_{\rm b}/\Omega_0}{0.17}}^{-1}
\left( \lambda_{\rm ta} \over 0.05 \right)^2 r_{\rm ta},
\label{eq:rrot}
\end{equation}
where $r_{\rm ta}$ is the turn-around radius of the shell, and we
adopt a median for the spin parameter, $\lambda_{\rm ta}=0.05$
(Efstathiou \& Jones 1979; Barns \& Efstathiou 1987; Warren et
al. 1992). The above radius defines the scale below which the gas is
supposed to attain rotational support and the approximation of
spherical symmetry breaks down. The overdensity of the gas exceeds
$\sim 10^5$ when it reaches this radius. The simulation is halted when
the innermost gas shell has fallen below this radius and is regarded
as `collapsed'.  On the other hand, a dark matter shell falling to the
center is rebounced at the radius $r_{\rm rot}$ of the innermost gas
shell, to mimic shell crossings.
 
At the outer boundary, we assume that the gas is in pressure
equilibrium, i.e., the pressure outside the cloud $P_{\rm out}$ is
equal to that of the outermost gas shell. This is different from the
free boundary condition ($P_{\rm out}=0$) adopted in Papers I and II.
While the inner structure of a simulated cloud is hardly affected by
this difference, we adopt the former in this paper so that the cloud
envelope maintains to follow the Hubble expansion throughout its
evolution.

\section{Results}
\label{sec:results}

\subsection{Significance of radiative transfer and H$_2$ formation}
\label{sec:signif}

\begin{figure*}
\begin{center}
   \leavevmode\psfig{figure=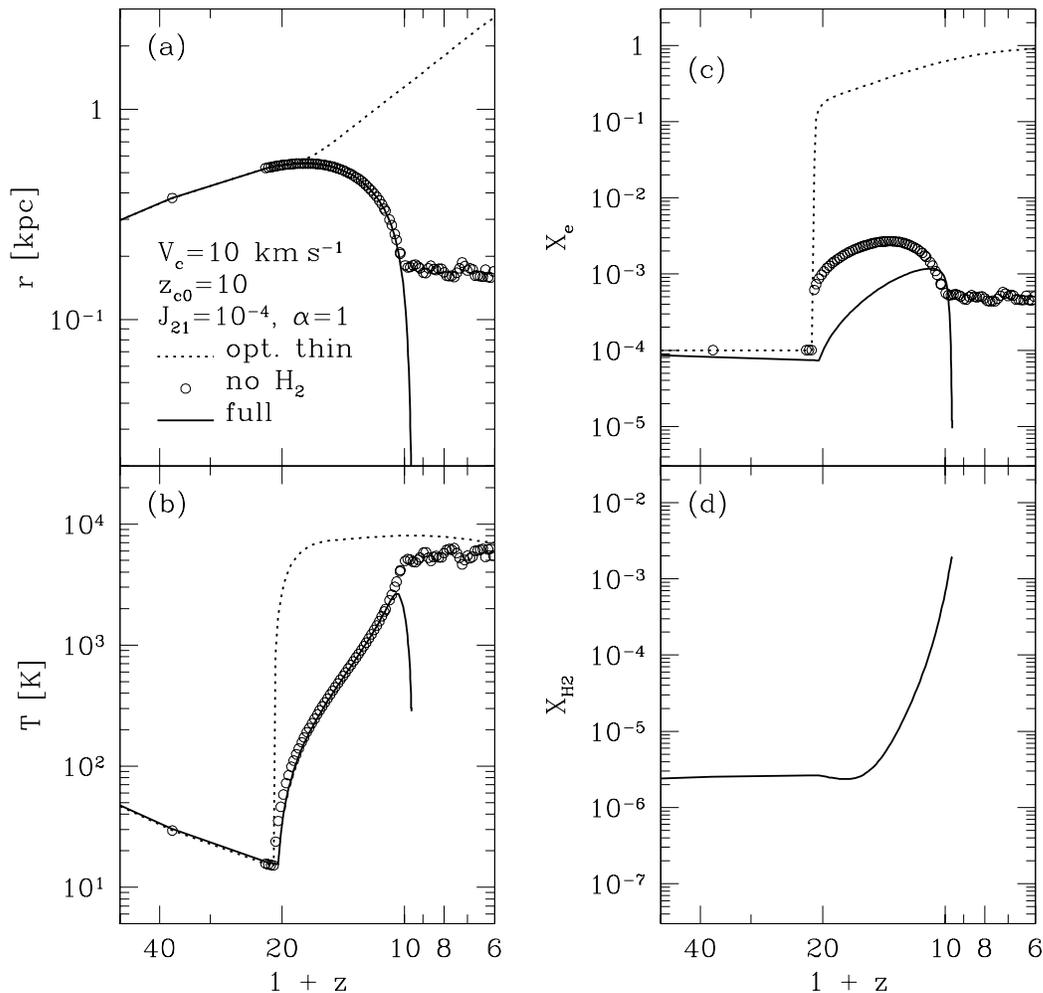,width=14cm}
\end{center}
\caption{Time evolution of (a) radius, (b) temperature, (c) electron
fraction, and (d) H$_2$ fraction, of the central gas shell of a low
mass object with $V_{\rm c}=10$ km s$^{-1}$ and 
$z_{\rm c0}=10$ ($z_{\rm c}=3$), comparing different 
treatments of radiative processes.  Lines and
symbols indicate results of the optically thin (dotted line), no H$_2$
(circles), and full (solid line) calculations with $J_{21}=10^{-4}$
and $\alpha=1$.}
\label{fig:rtxl}
\end{figure*}
The results of the full calculations are first compared with those of
previous approaches, to illustrate the significance of incorporating
radiative transfer and H$_2$ formation in our analyses. The most
noticeable changes are in the evolution of a low mass object with
$T_{\rm vir} < 10^4$K.  Fig. \ref{fig:rtxl} shows the time evolution
of the central part of a cloud with $z_{\rm \rm c0}=10$ ($z_{\rm c}=3$)
and $V_{\rm c}=10$ km s$^{-1}$, corresponding to $T_{\rm vir} = 4
\times 10^3$K. Both the dynamics and the thermal state are shown to
change completely under different treatments of the radiative
processes.  In the optically thin calculation, the cloud is
photoionized to the center and evaporated promptly by a little
radiation with $J_{21}=10^{-4}$ and $\alpha=1$. In the no H$_2$
calculation, the external radiation is attenuated by absorption, but
contraction of the gas is halted by thermal pressure as it lacks
coolant at $T < 10^4$K. In the full calculation, on the other hand,
the cloud center is kept self-shielded against the UVB and H$_2$
molecules in excess of $X_{\rm H2}=10^{-3}$ are produced, allowing the
gas to collapse and cool efficiently to $\sim 10^2$K. Details of the
H$_2$ formation and destruction processes in the UVB will be discussed
in Section \ref{sec:dyn}.

\begin{figure*}
\begin{center}
   \leavevmode\psfig{figure=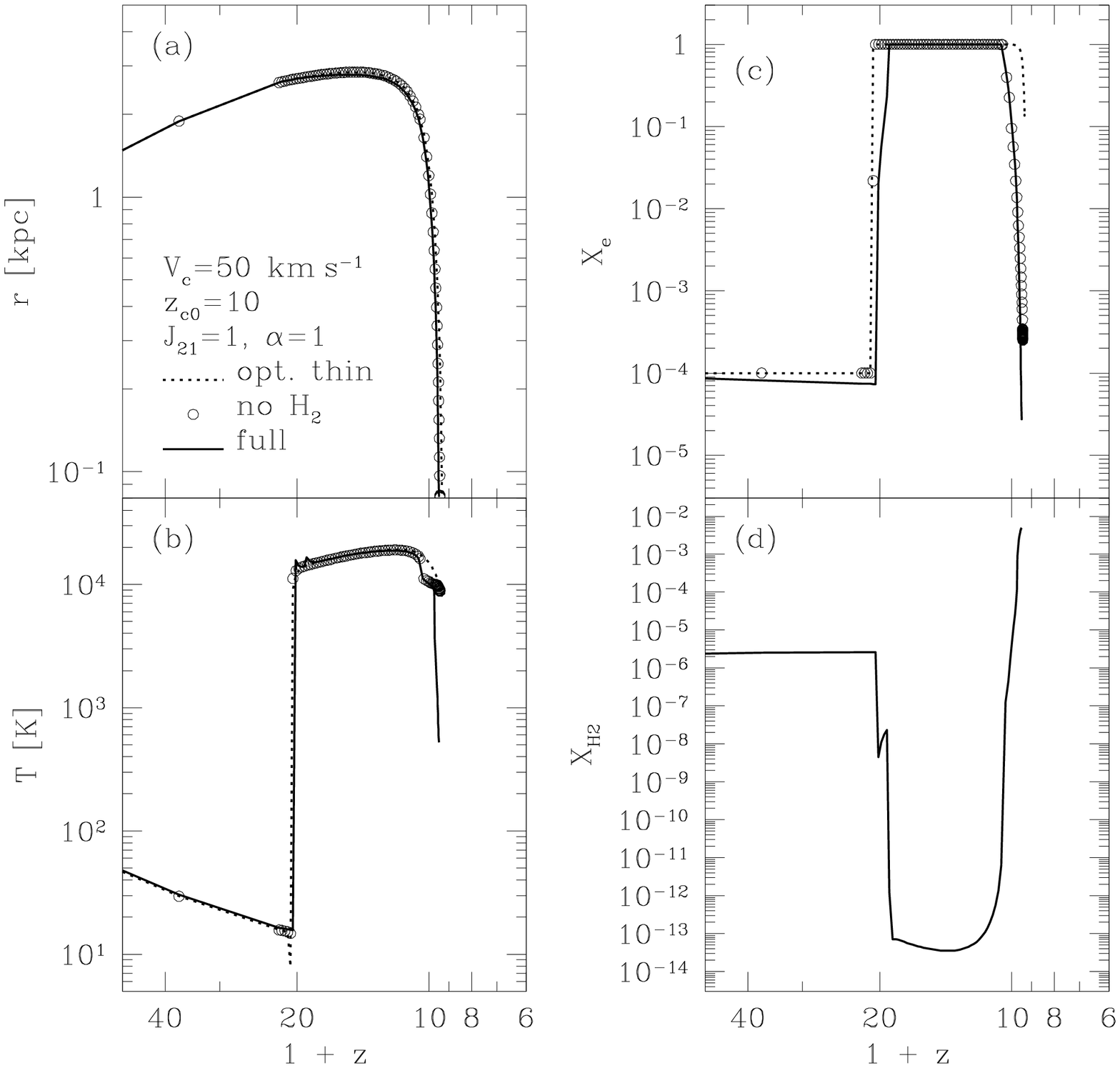,width=14cm}
\end{center}
\caption{Same as Fig.\protect\ref{fig:rtxl}\protect, except for
showing a high mass collapse with $V_{\rm c}=50$ km s$^{-1}$ in the UVB
with $J_{21}=1$ and $\alpha=1$.}
\label{fig:rtxh}
\end{figure*}
For a high mass object well above the scale $T_{\rm vir}=10^4$K, in
contrast, radiative transfer and H$_2$ formation appear to have
smaller impacts as far as the dynamics is concerned.
Fig. \ref{fig:rtxh} shows that such a cloud can collapse even in a
stronger UV field with $J_{21}=1$ and $\alpha=1$, regardless of
treatments of radiative processes. This is because gravitational
potential of the cloud is much greater than thermal energy attained by
photoionization ($T\sim 10^4$K) and atomic cooling is efficient enough
to enable the collapse.

\begin{figure*}
\begin{center}
   \leavevmode\psfig{figure=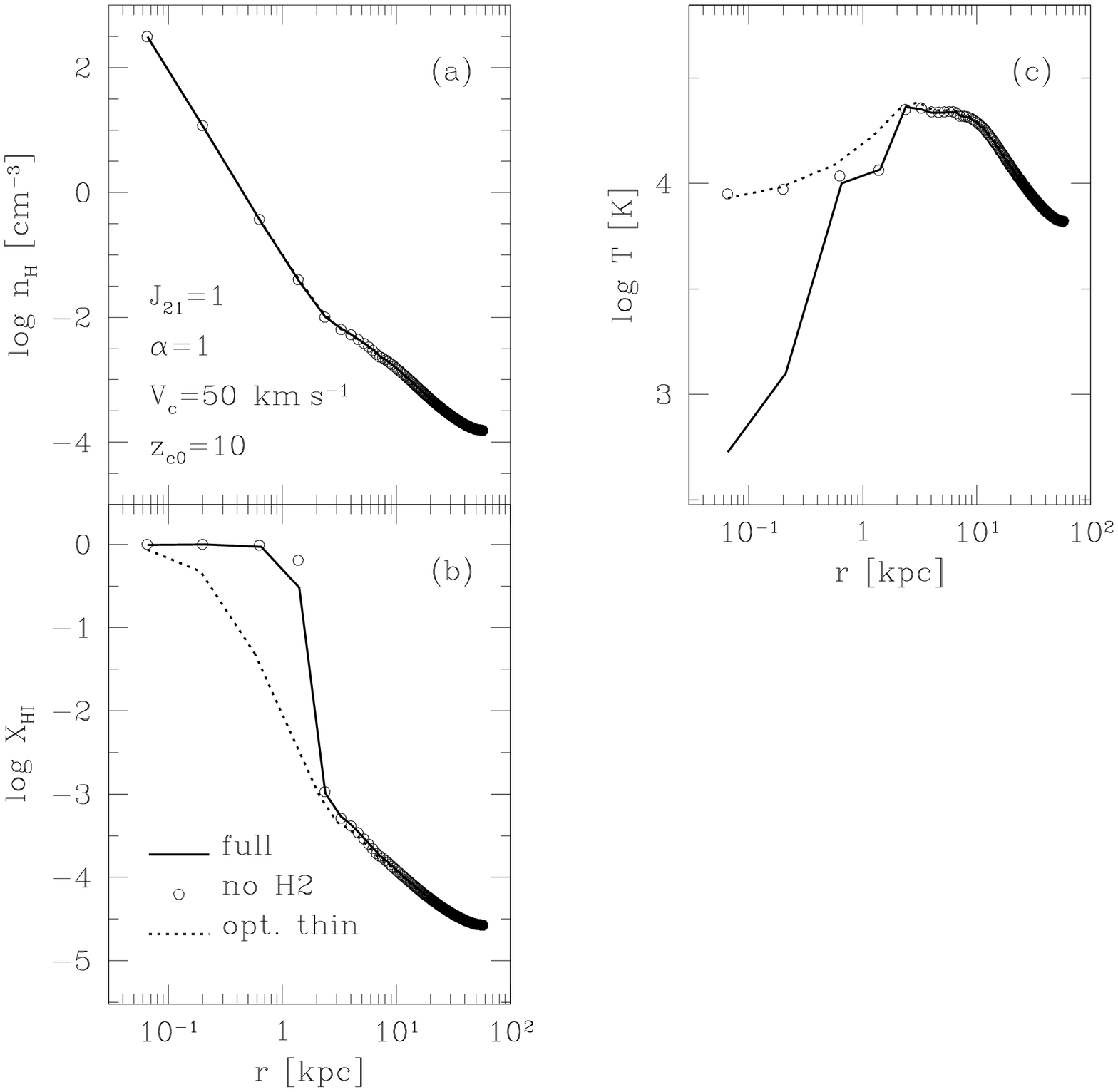,width=14cm}
\end{center}
\caption{Radial profiles at the collapse of the cloud center,
comparing the optically thin (dotted line), no H$_2$ (circles), and
full (solid line) calculations: (a) total hydrogen density, (b) HI
fraction, and (c) temperature.  $J_{21}=1$, $\alpha=1$, 
$V_{\rm c}=50$ km s$^{-1}$ and $z_{\rm c0}=10$ ($z_{\rm c}=3$) are adopted. }
\label{fig:prof1}
\end{figure*}
Nevertheless, the final thermal state of the gas is still modified to
a great extent in a high mass collapse, as indicated by the
differences in the temperature and electron fraction at the collapse
in Fig. \ref{fig:rtxh}. To see this point more clearly,
Fig. \ref{fig:prof1} compares the internal structure of the same
objects at the collapse of the central gas.  The density profile
commonly exhibits that the central gas is undergoing run-away collapse
(panel a), as it cools efficiently and contracts nearly at the
free-fall rate under the dark matter potential. The radiative transfer
effects, however, produce a self-shielded neutral region at $r \simlt
2$kpc (panel b), whereas $X_{\rm HI}$ is simply determined by
ionization equilibrium with uniform radiation in the optically thin
case.  The gas well inside this neutral region can cool down to $\sim
10^2$K once H$_2$ cooling is taken into account (panel c). In the
absence of H$_2$ molecules, the gas can cool only to $\sim 10^4$K by
atomic cooling.  Rapid increase of density due to run-away collapse is
in fact essential for self-shielding and H$_2$ cooling, as will be
discussed throughly in the forthcoming sections.

\begin{figure*}
\begin{center}
   \leavevmode\psfig{figure=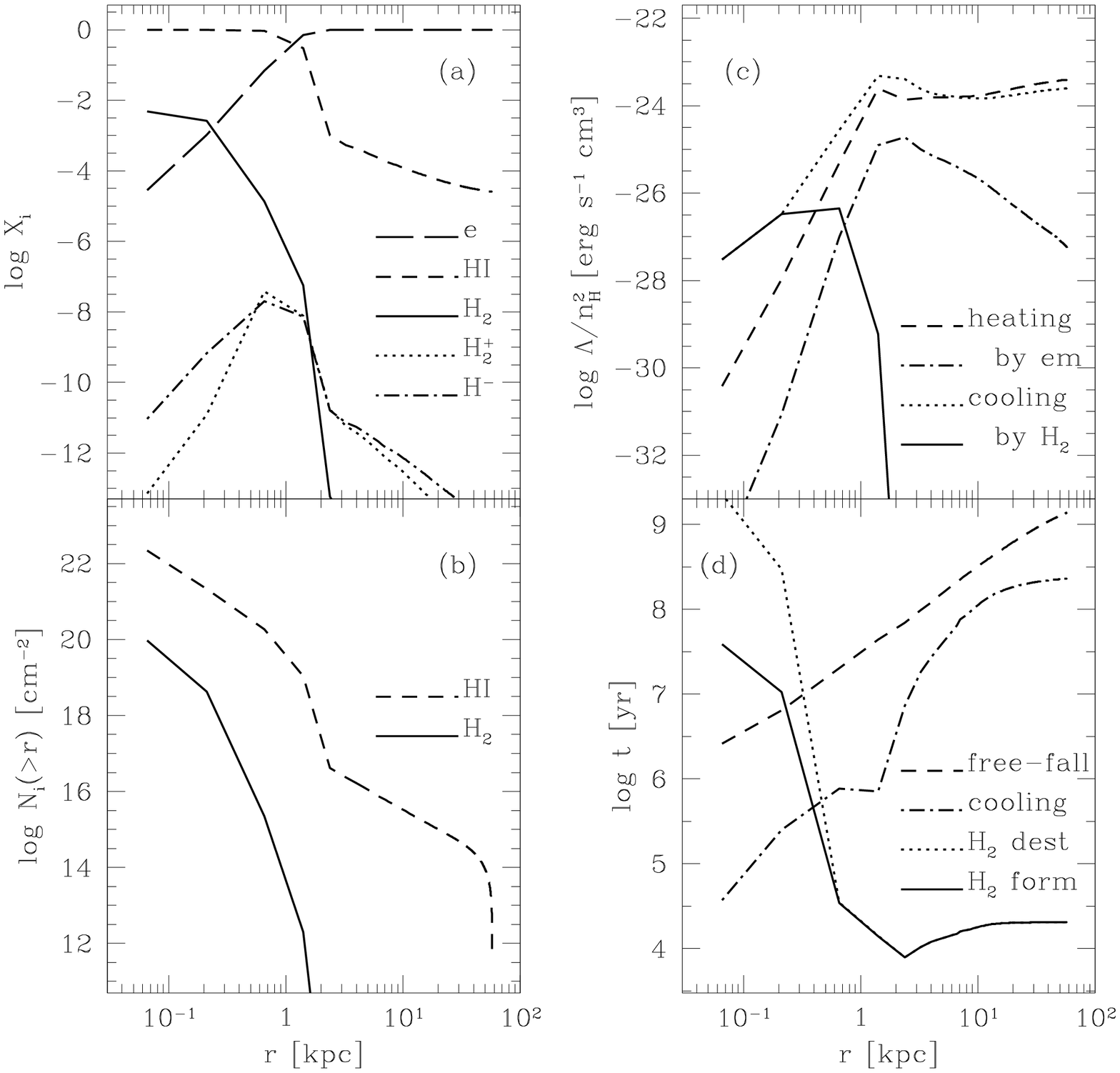,width=14cm}
\end{center}
\caption{Detailed radial profiles in the same full calculation as in
Fig.  \protect\ref{fig:prof1}\protect: (a) fractions of e (long
dashed), HI (short dashed), H$_2$ (solid), H$_2^+$ (dotted), and H$^-$
(dot-dashed), (b) column densities of HI (dashed) and H$_2$ (solid)
measured from the outer boundary, (c) rates of total UV heating
(dashed), UV heating by emitted photons (dot-dashed), total cooling
(dotted), and H$_2$ cooling (solid), and (d) timescales of free-fall
(dashed), cooling (dot-dashed), H$_2$ destruction (dotted), and H$_2$
formation (solid). }
\label{fig:prof2}
\end{figure*}
To further corroborate the above results, Fig. \ref{fig:prof2} shows
more detailed structures in the full calculation. While H$_2$
molecules are destroyed in the ionized outer envelope, they form
efficiently in the self-shielded central region (panel a). Roughly
speaking, the gas becomes optically-thick against ionizing ($>
13.6$eV) photons at $N_{\rm HI} \simgt 10^{17}$ cm$^{-2}$ and the LW
($11.2-13.6$eV) photons at $N_{\rm H2} \simgt 10^{14}$ cm$^{-2}$,
where $N_{\rm HI}$ and $N_{\rm H2}$ are respectively the HI and H$_2$
column densities measured from the outer boundary.  Both of these
conditions are amply satisfied at $r \simlt 1$ kpc (panel b). The UV
heating rate is reduced significantly in this optically thick region
(panel c). Photons emitted in the cloud interior contribute at most
$\sim 10\%$ to the total heating rate for the $\alpha=1$ spectrum,
though their contribution can be much larger for a softer spectrum
(Tajiri \& Umemura 2000). Cooling to temperatures below $10^4$K is
possible once H$_2$ cooling overtakes UV heating. At $T\simlt 10^4$K
($r \simlt 0.8$ kpc), the timescale of H$_2$ formation becomes shorter
than that of destruction as recombination lags behind the temperature
decrease (Kang \& Shapiro 1992; Corbelli, Galli \& Palla 1997), and
chemical reactions are out of equilibrium (panel d). The cooling
timescale, however, becomes even shorter as temperature drops, and
$X_{\rm H2}$ is frozen out at $\sim 10^{-3}$ at the cloud center. This
abundance is comparable to that achieved commonly in a metal-deficient
post-shock layer in the absence of radiation (Shapiro \& Kang 1987;
Ferrara 1998; Susa et al. 1998).

If helium is further incorporated into the present calculations, the
largest impact will be via heating by photons with $h \nu \geq 24.6$
eV. For given UV intensity $J_{21}$, HI number density $n_{\rm HI}$,
and HI column density $N_{\rm HI}$, the heating rate will be enhanced
by the addition of helium, in the case of $\alpha=1$, as
\begin{equation}
\frac{\cal H({\rm H+He})}{\cal H({\rm H})} \simeq
 \left\{\begin{array}{ll} 1 + 0.19 \sbkt{\frac{n_{\rm HeI}/n_{\rm
 HI}}{0.08}} & \mbox{optically thin limit}, \\ 1.46 \sbkt{\frac{n_{\rm
 HeI}/n_{\rm HI}}{N_{\rm HeI}/N_{\rm HI}}} & \mbox {optically thick
 limit},\\
\end{array} \right. 
\label{eq:he}
\end{equation}
where ${\cal H}({\rm H+He})$ is the total heating rate in units of erg
s$^{-1}$ cm$^{-3}$ of both H and He, and ${\cal H}({\rm H})$ is that
of the pure hydrogen gas. These rates are computed as in the Appendix
of Paper I, which are in good agreement with more rigorous
treatment. Equation (\ref{eq:he}) indicates that the total heating
rate tends to increase by $\sim 20 - 50\%$ once helium with the
primordial abundance ($\sim 8\%$ by number) is taken into account.
This is much smaller than the overall variations in the heating and
cooling rates in a cloud shown in Fig. \ref{fig:prof2}(c). In fact, we
have checked that an additional heat input of $50\%$ does not
considerably alter either the time evolutions in Figs \ref{fig:rtxl}
and \ref{fig:rtxh} or the internal structures in Figs \ref{fig:prof1}
and \ref{fig:prof2}.  Though temperature of the diffuse outer envelope
can increase by as much as a factor of 2 by helium heating (Abel \&
Haehnelt 1999), the evolution of the central high-density region, in
which our main interests reside, is little influenced. Helium heating
will be of less importance for softer spectra. Hence we neglect helium
in our analyses.

In summary, the above results justify our present approach and confirm
the importance of including radiative transfer of UV photons, H$_2$
formation/destruction, and hydrodynamics, in studying the development
of population III objects in the UVB.

\subsection{Dynamical and thermal evolution}
\label{sec:dyn} 

\begin{figure*}
\begin{center}
   \leavevmode\psfig{figure=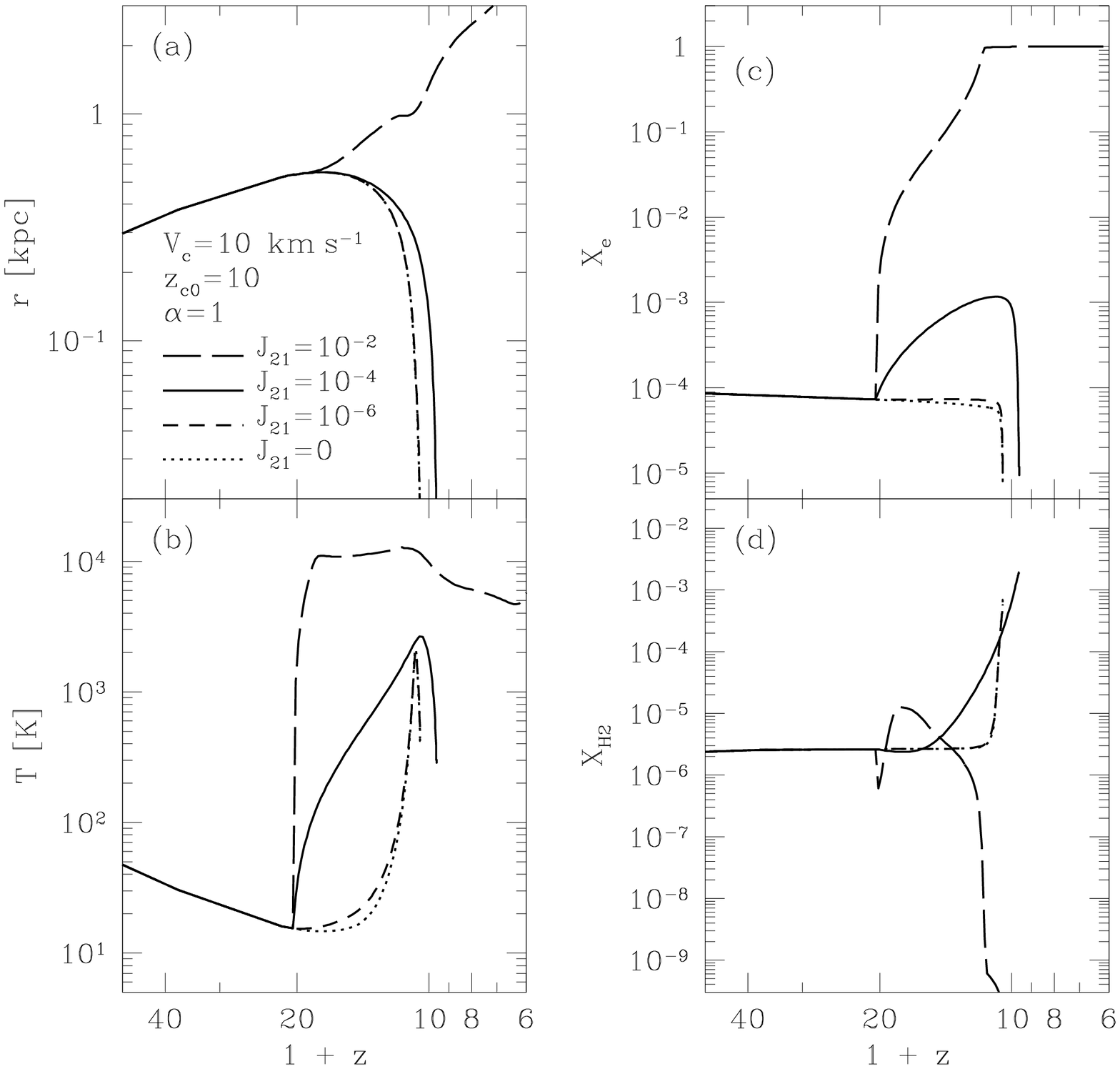,width=14cm}
\end{center}
\caption{Same as Fig.\protect\ref{fig:rtxl}\protect, except for
comparing different values of $J_{21}$ in the case of
$\alpha=1$. Lines indicate the results with $J_{21}=10^{-2}$ (long
dashed), $10^{-4}$ (solid), $10^{-6}$ (short dashed), and $0$
(dotted). The results of $J_{21}=10^{-6}$ and 0 closely overlap each
other.}
\label{fig:rtxj}
\end{figure*}
The evolution of UV irradiated clouds are studied in more detail in
the following. Fig. \ref{fig:rtxj} displays how the time evolution of
a low mass object ($V_{\rm c}=10$ km s$^{-1}$) depends on the radiation
intensity. In the absence of the UVB ($J_{21}=0$), the cloud decoupled
from the Hubble expansion is first shock-heated and then cools
radiatively. At $J_{21}=10^{-2}$, in contrast, the gas is photoionized
and evaporated even if radiative transfer is taken into account. Once
the gas is photoionized and heated, H$_2$ molecules are destroyed by a
large amount. At lower intensity ($J_{21}=10^{-4}$), as was also shown
in Fig. \ref{fig:rtxl}, the cloud can collapse and cool down by H$_2$.
At even weaker intensity ($J_{21}=10^{-6}$), the radiation is
attenuated almost completely and the evolution resembles that in the
no UV case. Note that the H$_2$ fraction increases after the onset of
the UVB at $J_{21}=10^{-4}$, compared to the no UV case. This is
indeed the case in which the formation of H$_2$ via H$^-$ and H$_2^+$
is promoted by an enhanced ionized fraction in a weakly photoionized
medium (Haiman et al. 1996b).

\begin{figure*}
\begin{center}
   \leavevmode\psfig{figure=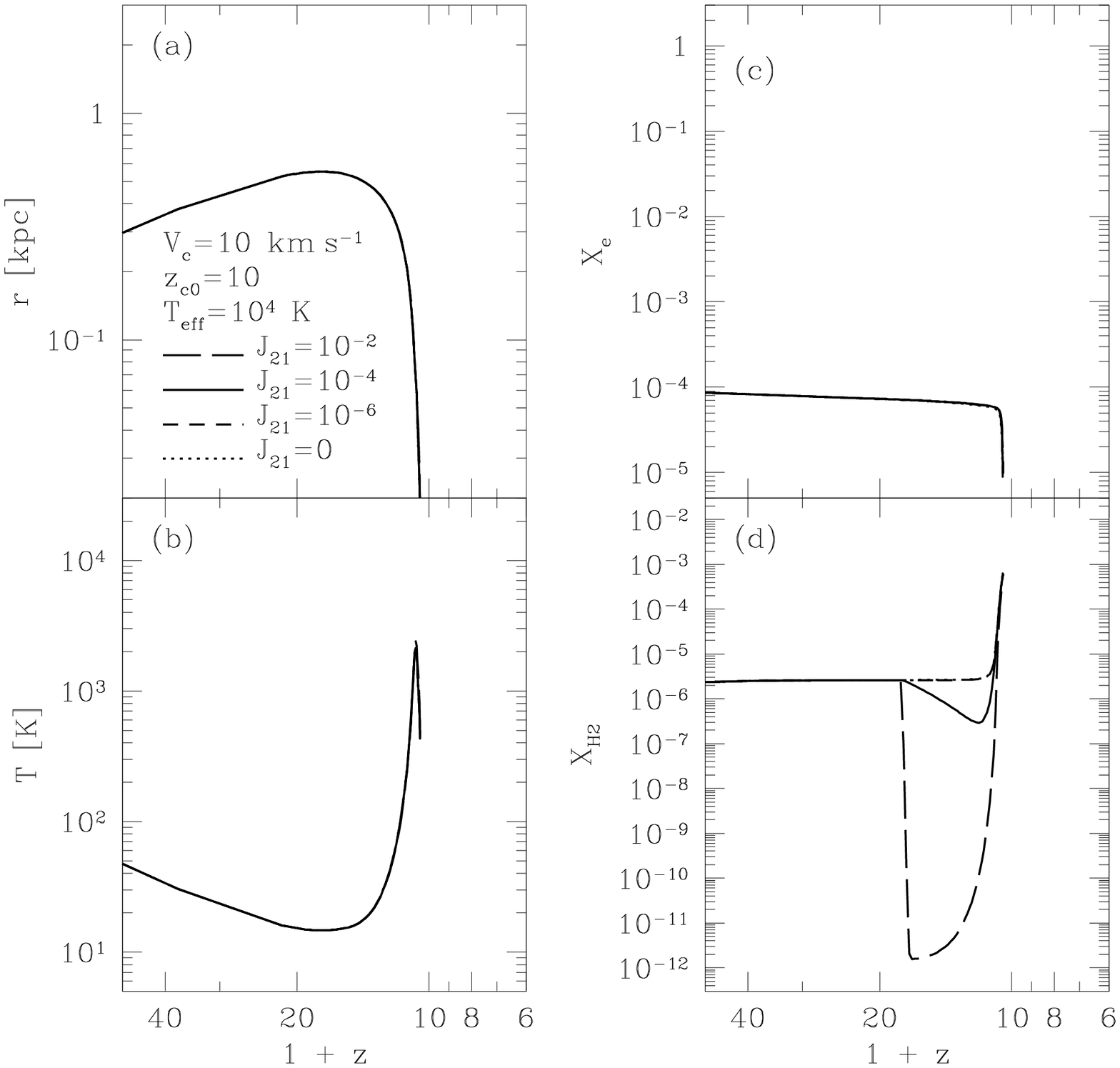,width=14cm}
\end{center}
\caption{Same as Fig.\protect\ref{fig:rtxj}\protect, except for
adopting a black-body spectrum with $T_{\rm eff}=10^4$K. In panels
(a)--(c), all the lines overlap one another.}
\label{fig:rtxaj}
\end{figure*}
Fig. \ref{fig:rtxaj} illustrates that the shape of the radiation
spectrum also influences greatly the thermal and dynamical evolution.
Compared to the $\alpha=1$ case, the black-body spectrum with $T_{\rm
eff}=10^4$K is softer and has much less ionizing photons with energies
above $13.6$eV (see Fig. 1 and Table 1). For the same values of
$J_{21}$ as in Fig. \ref{fig:rtxj}, the black-body UVB therefore has
weaker impacts on the evolution of radius, temperature, and the
ionization degree. On the other hand, H$_2$ molecules are destroyed by
a larger amount at the onset of the UVB, because a greater number of
photons with energies below $13.6$eV are available to destroy H$_2$
and its intermediates, H$^-$ and H$_2^+$. Nevertheless, as the gas
contracts, it becomes optically thick against the LW photons and the
final H$_2$ fraction of $X_{\rm H2} \sim 10^{-3}$ is still achieved.

\begin{figure*}
\begin{center}
   \leavevmode\psfig{figure=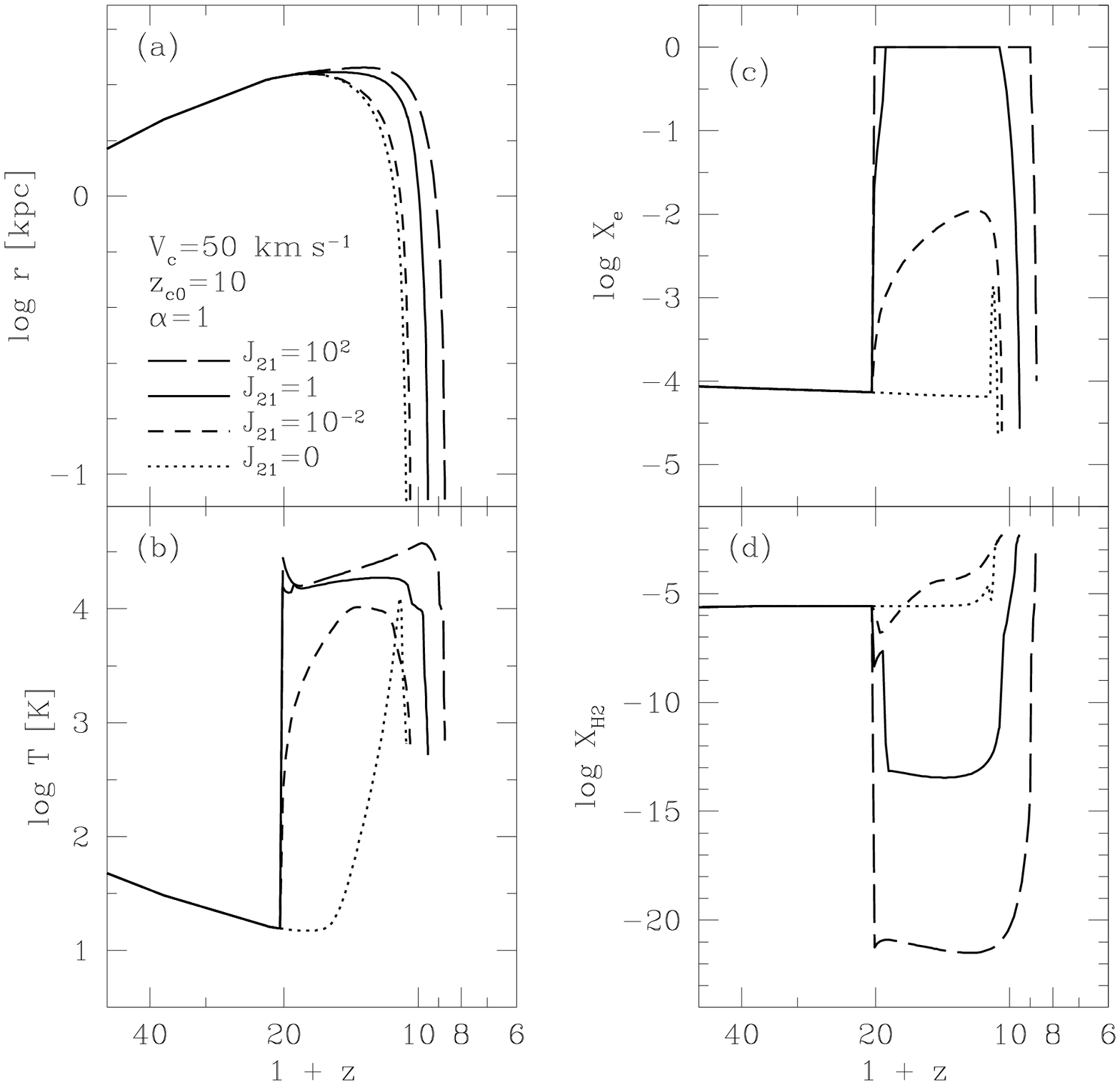,width=14cm}
\end{center}
\caption{Same as Fig.\protect\ref{fig:rtxj}\protect, except for
showing a high mass collapse with $V_{\rm c}=50$ km s$^{-1}$ in the UVB
with $J_{21}=10^{2}$ (long dashed), $1$ (solid), $10^{-2}$ (short
dashed), and $0$ (dotted).}
\label{fig:rtxhj}
\end{figure*}
For a high mass object well above the scale $T_{\rm vir}=10^4$K, the
collapse is hardly suppressed even under the strong UVB, but its
thermal evolution remains sensitive to the UVB.  As shown in
Fig. \ref{fig:rtxhj}, for $J_{21}\simgt 1$ and $\alpha=1$, the gas is
promptly heated to $\sim 10^4$K and H$_2$ molecules are destroyed
mainly via collisions with H$^+$. At $J_{21}=10^{-2}$, on the other
hand, the H$_2$ abundance is enhanced compared to the no UV case by an
increased degree of ionization, as already mentioned.  In either case,
the H$_2$ fraction at the collapse again exceeds $X_{\rm H2} \sim
10^{-3}$.

\begin{figure*}
\begin{center}
   \leavevmode\psfig{figure=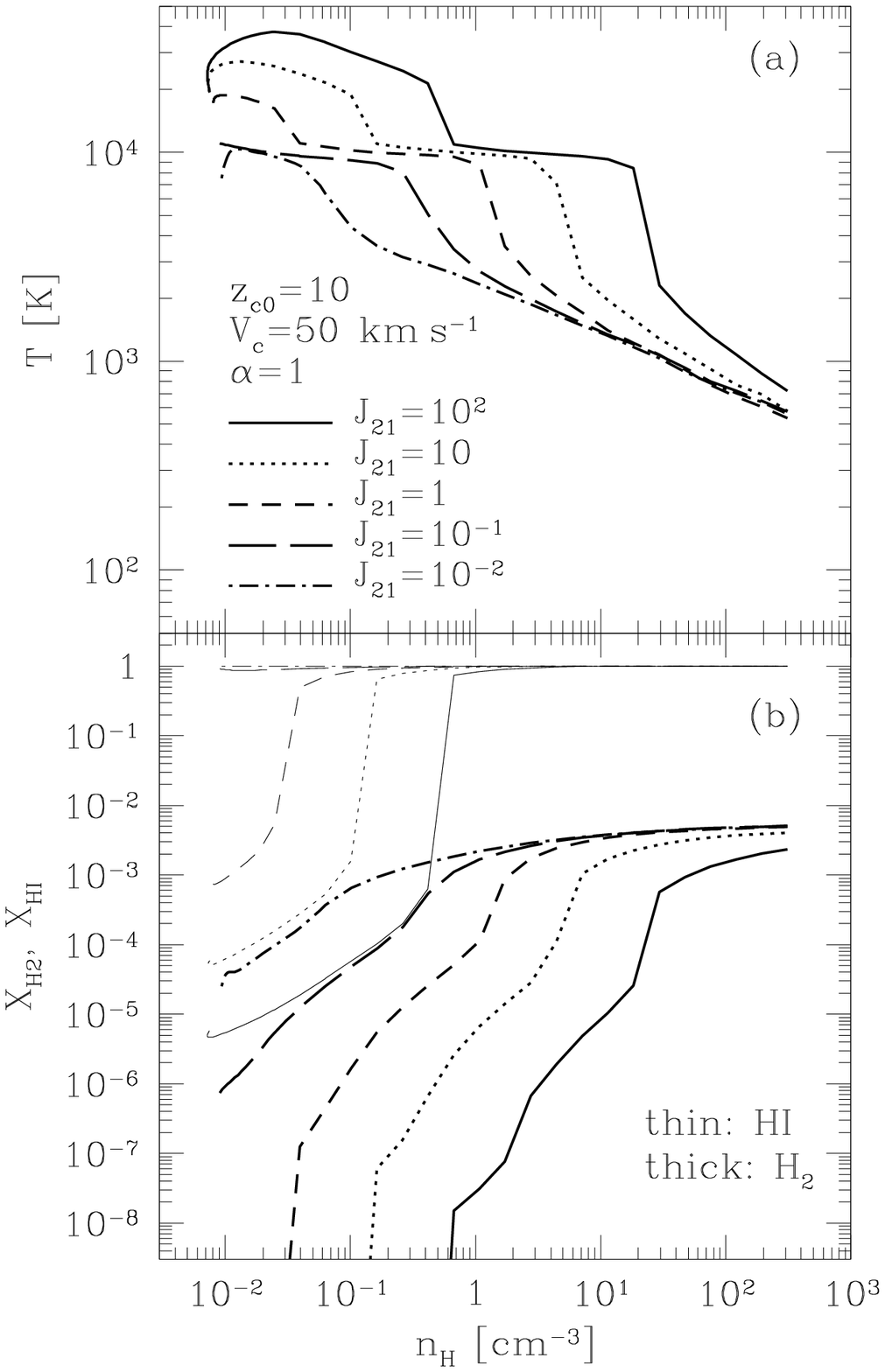,width=14cm}
\end{center}
\caption{Evolution of (a) temperature, (b) $X_{\rm H2}$ (thick lines)
and $X_{\rm HI}$ (thin lines), versus gas density at the center of a
collapsing cloud with $V_{\rm c}=50$ km s$^{-1}$ and $z_{\rm c0}=10$ ($z_{\rm
c}=3$) in the case of $\alpha=1$. Lines indicate $J_{21}=10^2$
(solid), 10 (dotted), 1 (short dashed), $10^{-1}$ (long dashed), and
$10^{-2}$ (dot dashed). For clarity, only the evolution at $z<15$ is
plotted.}
\label{fig:core1}
\end{figure*}
\begin{figure*}
\begin{center}
   \leavevmode\psfig{figure=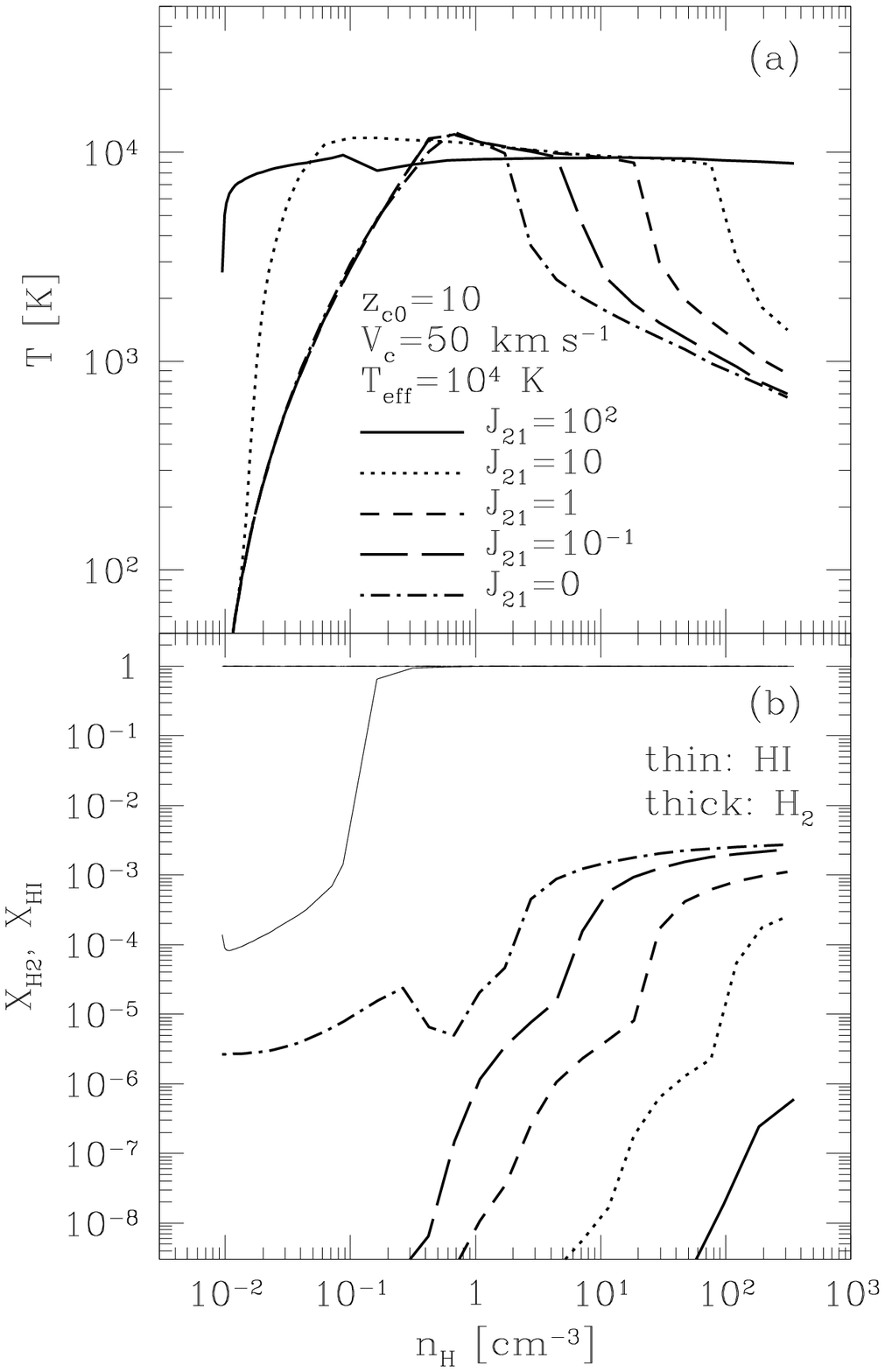,width=14cm}
\end{center}
\caption{Same as Fig.\protect\ref{fig:core1}\protect, except that
$T_{\rm eff}=10^4$K is adopted and dot-dashed lines indicate
$J_{21}=0$. In panel (b), thin lines ($X_{\rm HI}$) overlap one
another for $J_{21}\leq 10$.}
\label{fig:core2}
\end{figure*}
To illustrate more clearly how self-shielding and radiative cooling
operate in a collapsing cloud, we plot in Figs \ref{fig:core1} and
\ref{fig:core2} the temperature and fractions of HI and H$_2$ at the
cloud center as a function of its density. As the gas contracts and
density rises, its state changes rather suddenly from the hot ($T >
10^4$K) ionized `HII phase', to the warm ($T \sim 10^4$K) neutral
`HI phase', and the cold ($T < 10^4$K) neutral `H$_2$ phase'.
Such transitions are analogous to those pointed out by Kepner et
al. (1997), except that they are induced here by the dynamical
contraction of the gas itself rather than the decline of the external
UV intensity. The transition from the HII phase to the HI one takes
place when the cloud becomes optically thick against ionizing photons. 
The gas finally enters the H$_2$ phase once the cooling timescale
becomes shorter than the other relevant timescales, such as those of
free-fall and H$_2$ formation/destruction.  These figures confirm that
$X_{\rm H2}$ tends to converge to $\sim 10^{-3}$ after the H$_2$ phase
is achieved.

\begin{figure*}
\begin{center}
   \leavevmode\psfig{figure=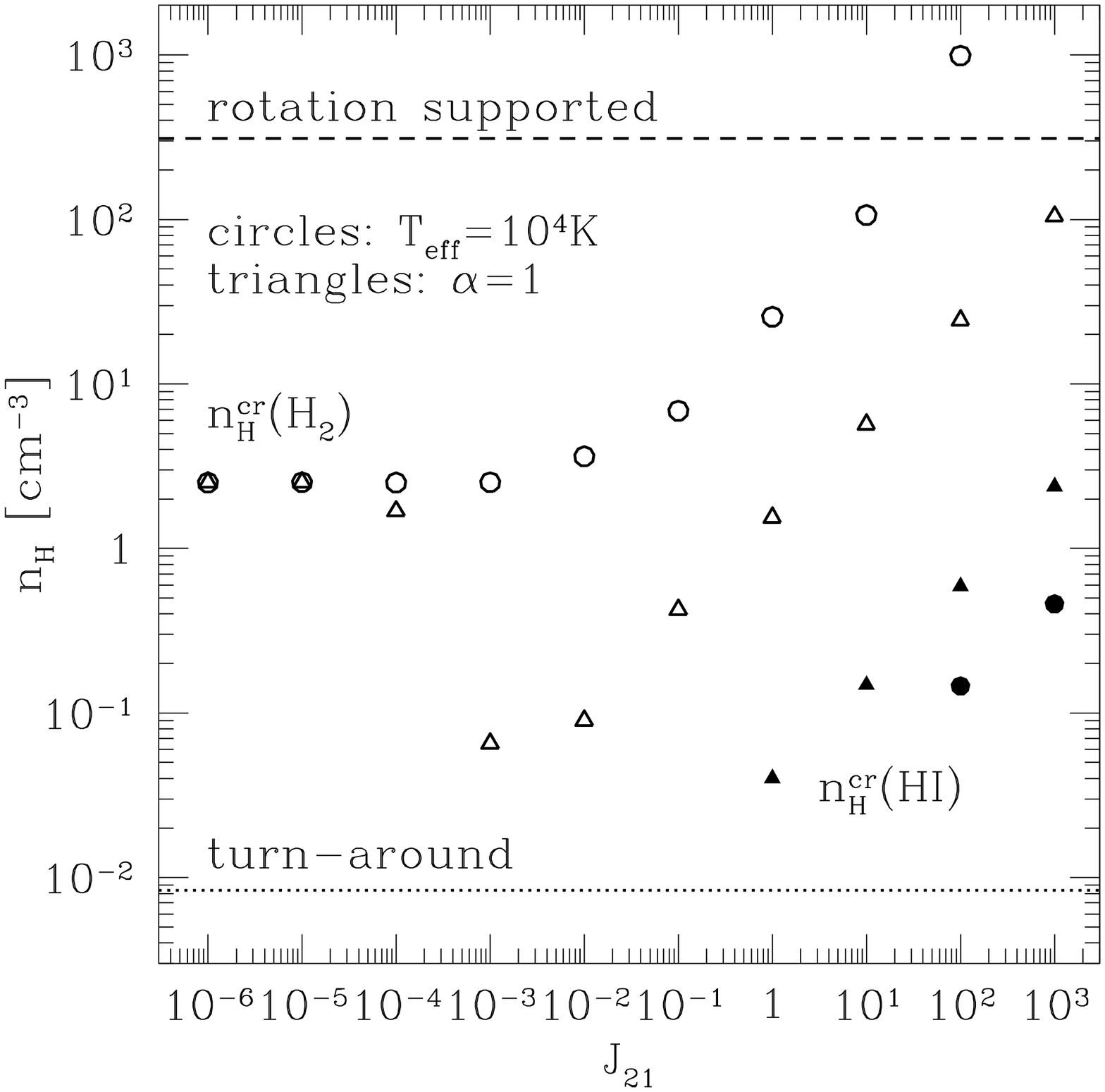,width=10cm}
\end{center}
\caption{Critical central densities for HI shielding $n_{\rm H}^{\rm
cr}(\mbox{HI})$ ($X_{\rm HI} > 0.5$, filled symbols) and H$_2$ cooling
$n_{\rm H}^{\rm cr}(\mbox{H$_2$})$ ($T < 5000$K, open symbols) versus
$J_{21}$.  The spectral shape of the UVB is specified by $\alpha=1$
(triangles) and $T_{\rm eff}=10^4$K (circles).  Also plotted for
reference are the densities at which the cloud center turns around
(dotted line) and is rotationally supported (dashed line). $V_{\rm c}=50$ km
s$^{-1}$ and $z_{\rm c0}=10$ ($z_{\rm c}=3$) are adopted.}
\label{fig:nj}
\end{figure*}
Figs \ref{fig:core1} and \ref{fig:core2} further indicate that
quantitative details of the transitions among the above phases depend
sensitively on the intensity and spectrum of the UVB. To examine this
more directly, we plot in Fig. \ref{fig:nj} the critical densities for
HI shielding $n_{\rm H}^{\rm cr}(\mbox{HI})$ and H$_2$ cooling $n_{\rm
H}^{\rm cr}(\mbox{H$_2$})$, defined as those at which the central HI
fraction exceeds 0.5 and temperature falls below 5000K,
respectively. In what follows, we regard the gas as in the HII, HI,
and H$_2$ phases, if the density is in the range $n_{\rm H} < n_{\rm
H}^{\rm cr}(\mbox{HI})$, $n_{\rm H}^{\rm cr}(\mbox{HI}) < n_{\rm H} <
n_{\rm H}^{\rm cr}(\mbox{H$_2$})$, and $n_{\rm H} > n_{\rm H}^{\rm
cr}(\mbox{H$_2$})$, respectively.  Though Fig. \ref{fig:nj} is plotted
for particular choices of $V_{\rm c}$ and $z_{\rm c}$, basic features
are similar in other cases. A possible change is that for smaller
masses and large $J_{21}$ the collapse is completely prohibited (see
Fig. \ref{fig:jv}) and the critical densities simply disappear.

Fig. \ref{fig:nj} shows that $n_{\rm H}^{\rm cr}(\mbox{HI})$ increases
with $J_{21}$ for a given spectral shape, i.e., the higher density is
required to shield the stronger ionizing flux. At small $J_{21}$,
$n_{\rm H}^{\rm cr}(\mbox{HI})$ falls below the turn-around density,
indicating that photoionization is kept attenuated throughout the
cloud evolution. This feature is also apparent in Figs \ref{fig:core1}
and \ref{fig:core2}, in which $X_{\rm HI}$ is kept high at weak UV
intensities. Compared to the $\alpha=1$ case, the $T_{\rm eff}=10^4$K
spectrum yields lower values of $n_{\rm H}^{\rm cr}(\mbox{HI})$,
because there are less ionizing ($>13.6$ eV) photons for a given
$J_{21}$. 

On the other hand, $n_{\rm H}^{\rm cr}(\mbox{H$_2$})$ behaves in a
more complex manner, depending on the spectral shape. For $\alpha=1$,
it is not a monotonic function of $J_{21}$, but has a minimum at
moderate intensities $J_{21}\sim 10^{-3}-10^{-2}$. Formation of H$_2$
molecules is enhanced maximally at these intensities. At weaker
intensities $J_{21}\simlt 10^{-4}$ the result almost coincides with
that of the no UV case, whereas at stronger intensities $J_{21}\simgt
10^{-1}$ H$_2$ formation is suppressed to higher densities.  For
$T_{\rm eff}=10^4$K, in contrast, a greater fraction of photons below
13.6 eV is present to suppress H$_2$ formation. The threshold $n_{\rm
H}^{\rm cr}(\mbox{H$_2$})$ is therefore higher than that of $\alpha=1$
for a given $J_{21}$.  This is the reason why the convergence of
$X_{\rm H2}$ is rather slow in Fig. \ref{fig:core2}.  In either case,
Fig. \ref{fig:nj} assures that the H$_2$ phase is realized at the
center of a contracting cloud before it is rotationally supported, as
long as $J_{21}$ is not far above the level $J_{21} \sim 1$ observed
at $z\sim 3$.

The above critical densities $n_{\rm H}^{\rm cr}(\mbox{HI})$ and
$n_{\rm H}^{\rm cr}(\mbox{H$_2$})$ at $J_{21} \simgt 1$ with
$\alpha=1$ are in rough agreement with analytical estimations by
Tajiri \& Umemura (2000) for an isothermal sphere; $n_{\rm H}^{\rm
cr}(\mbox{HI}), ~n_{\rm H}^{\rm cr}(\mbox{H$_2$}) \propto
J_{21}^{2/3}$.  This confirms, in the case of $\alpha=1$, that $n_{\rm
H}^{\rm cr}(\mbox{HI})$ is essentially the density at which the number
of recombinations in a cloud exceeds that of ionizing photons.
Similarly, $n_{\rm H}^{\rm cr}(\mbox{H$_2$})$ at $J_{21} \simgt 1$
corresponds to the density at which the H$_2$ cooling rate overcomes
the UV heating rate attenuated at the cloud center. For a much softer
spectrum ($T_{\rm eff}=10^4$K), the deviations from such simplified
estimations become larger, because of the greater significance of
emitted photons and the LW photons.

\subsection{Criteria for collapse and H$_2$ cooling}
\label{sec:thre}

We have performed the simulations for a variety of parameters, such as
the intensity and spectrum of the UVB, and the circular velocity and
collapse redshift of a cloud. The results are summarized in the
following.

\begin{figure*}
\begin{center}
   \leavevmode\psfig{figure=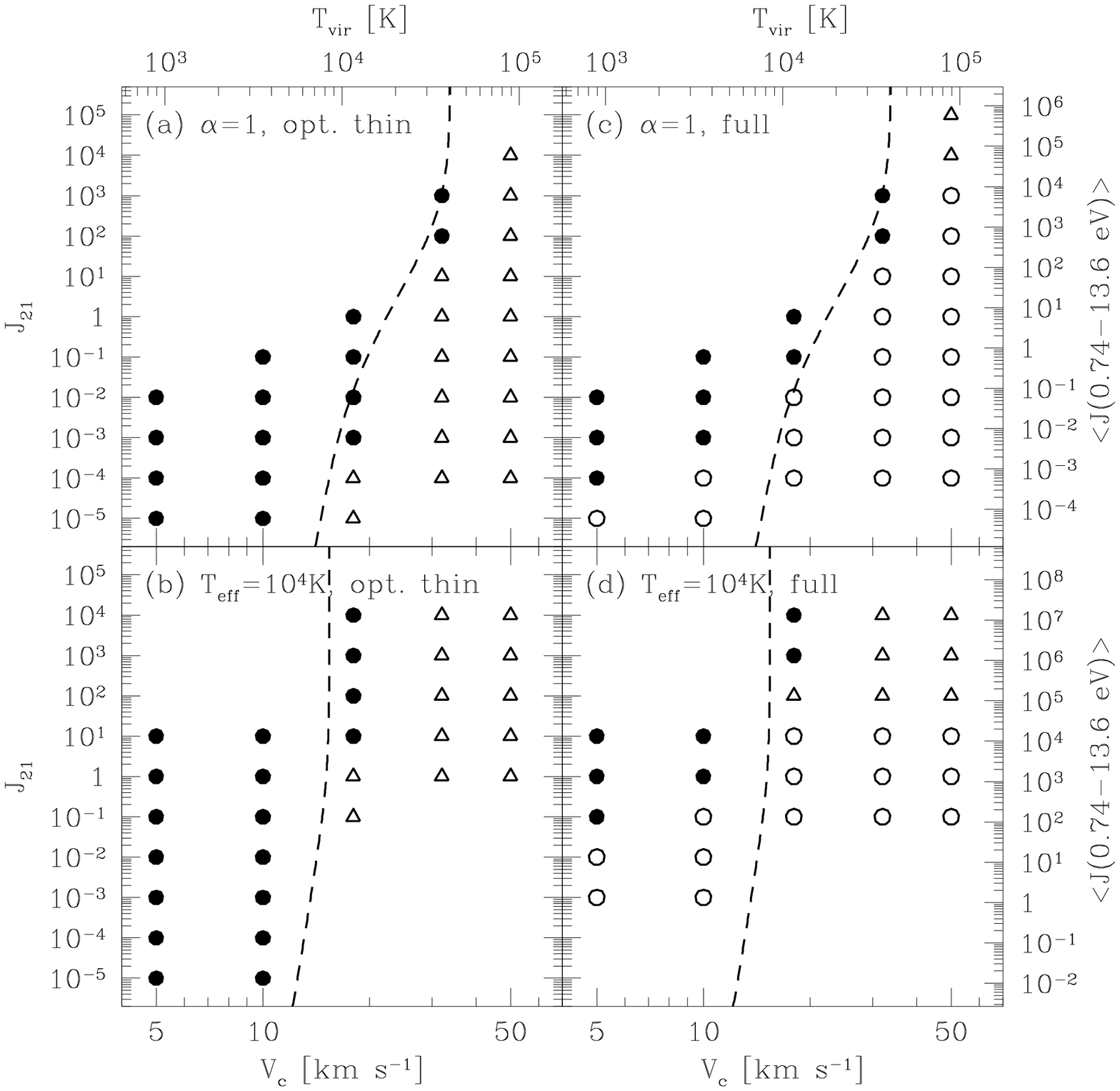,width=14cm}
\end{center}
\caption{$J_{21}- V_{\rm c}$ diagram on collapse and cooling of a
cloud with $z_{\rm \rm c0}=10$ ($z_{\rm c}=3$). Different panels
compare the optically thin (left) and full (right) calculations, and
the UV spectrum with $\alpha=1$ (top) and $T_{\rm eff}=10^4$K
(bottom).  Symbols indicate whether the cloud center collapses and the
H$_2$ phase is realized (open circles), it collapses but the H$_2$
phase is not realized (open triangles), or it is prohibited to
collapse (filled circles). Dashed lines show an analytical estimation
for the collapse threshold in the optically thin limit defined in
Paper II. The average intensity below the Lyman limit $\langle J
(\mbox{0.74--13.6 eV})\rangle$ is displayed on the right vertical
axis.}
\label{fig:jv}
\end{figure*}
Fig. \ref{fig:jv} shows a $J_{21} - V_{\rm c}$ diagram on collapse and
cooling of a cloud with $z_{\rm c0}=10$ ($z_{\rm c}=3$). The final
state of the central gas is classified into three categories: (i) it
collapses and cools to $T < 5000$K, i.e. the H$_2$ phase is realized;
(ii) it collapses but does not cool to $T< 5000$K; and (iii) it is
prohibited to collapse within the present age of the universe. Also
plotted for reference is the relation $T_{\rm vir}=T_{\rm eq}^{\rm
max}$, where $T_{\rm eq}^{\rm max}$ is the maximum equilibrium
temperature defined semi-analytically in Section 3.3 of Paper II. This
relation roughly corresponds to the Jeans condition at the maximum
expansion of the central gas and provides a reasonable estimation for
the collapse threshold in the optically thin limit.

The predictions of the full calculations are distinct from the
optically thin results in two ways. One is that the collapse of low
mass objects with $T_{\rm vir} < 10^4$K ($V_{\rm c} < 17$ km s$^{-1}$)
is possible at small UV intensities. The other is that the central gas
in collapsed clouds can cool efficiently to temperatures well below
$10^4$K, as long as $J_{21}$ is not far above the observed level
$J_{21} \sim 1$. The suppression of H$_2$ cooling occurs at
$J_{21}\simgt 10^4$ for $\alpha=1$ and $J_{21}\simgt 10^2$ for $T_{\rm
eff}=10^4$K, mainly due to photodetachment of H$^-$ as well as
photodissociation of H$_2$. To quantify this effect, we indicate in
Fig. \ref{fig:jv} the average intensity of the external UVB defined by
equation (\ref{eq:jb}) at 0.74--13.6 eV. The suppression of H$_2$
cooling is shown to take place at $\langle J (\mbox{0.74--13.6
eV})\rangle \simgt 5 \times 10^4$ in massive clouds for both spectra.

\begin{figure*}
\begin{center}
   \leavevmode\psfig{figure=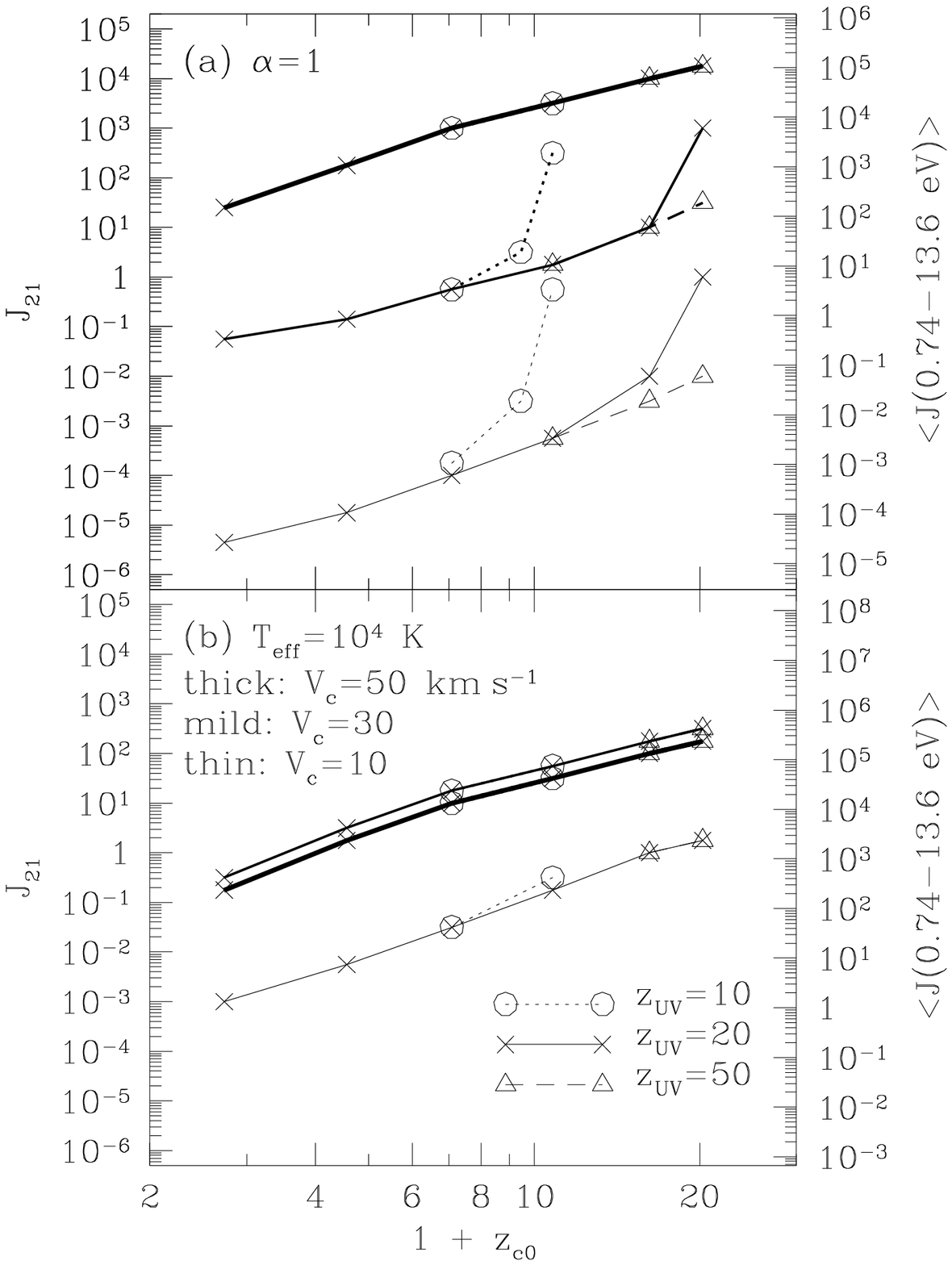,width=14cm}
\end{center}
\caption{Critical UV intensity above which H$_2$ cooling is prevented
as a function of central collapse redshift $z_{\rm c0}$: (a) $\alpha=1$,
and (b) $T_{\rm eff}=10^{4}$K.  Circular velocity of a cloud is taken
to be $V_{\rm c}=50$ (thick lines), 30 (mildly thick lines) and 10
km s$^{-1}$ (thin lines). The onset of the UVB is varied as $z_{\rm UV}=10$
(circles, dotted lines), 20 (crosses, solid lines) and 50 (triangles,
dashed lines).}
\label{fig:jcrit}
\end{figure*}

Given circular velocity and collapse redshift of a cloud, one can
compute the critical radiation intensity for collapse and H$_2$
cooling.  Fig. \ref{fig:jcrit} exhibits, as a function of the central
collapse redshift $z_{\rm c0}$, the intensity above which H$_2$
cooling is prevented completely in a cloud, i.e., the final state of
the central gas belongs to the category (ii) or (iii) defined above.
The critical intensity is in general an increasing function of
redshift, i.e., gas density. It also increases with circular velocity
but converges at $V_{\rm c} \simgt 50$ km s$^{-1}$ and $V_{\rm c}
\simgt 30$ km s$^{-1}$ for $\alpha=1$ and $T_{\rm eff}=10^{4}$K,
respectively. The convergence takes place once photoheating no longer
affects the cloud evolution and only photodestruction of H$_2$ and
H$^-$ becomes important. The converged intensities therefore have
nearly identical average values below the Lyman limit $\langle J
(\mbox{0.74--13.6 eV})\rangle \sim 30(1+z_{\rm c0})^3$ for both
spectra. Once converged, the critical intensities tend to drop
slightly for larger $V_{\rm c}$ (panel b) because the electron
fraction of the gas is lower in a cloud with higher hydrogen column
densities.

To explore the possibilities of different epochs of reionization, we
have added to Fig. \ref{fig:jcrit} the results of runs with $z_{\rm
UV}=10$ and 50. If $z_{\rm c0}$ is sufficiently close to $z_{\rm UV}$,
the central gas is already contracting when it is exposed to the UVB,
and hence the stronger intensity is required to prevent collapse and
cooling. The difference is greater for the harder spectrum and the
smaller cloud mass, for which suppression of cooling is mainly via
photoevaporation and depends highly on the initial dynamical state of
the gas. For the softer spectrum and the larger cloud mass, on the
other hand, the suppression is regulated to a greater extent by
photodestruction of H$_2$ and H$^-$, and is less sensitive to the
initial dynamical state. After all, Fig. \ref{fig:jcrit} indicates
that there is a greater chance for the gas to cool even in a cloud
well below the scale $V_{\rm c}=30$ km s$^{-1}$ at high redshifts,
provided that the intensity is comparable to or less than $J_{21}\sim
1$.

\begin{figure*}
\begin{center}
   \leavevmode\psfig{figure=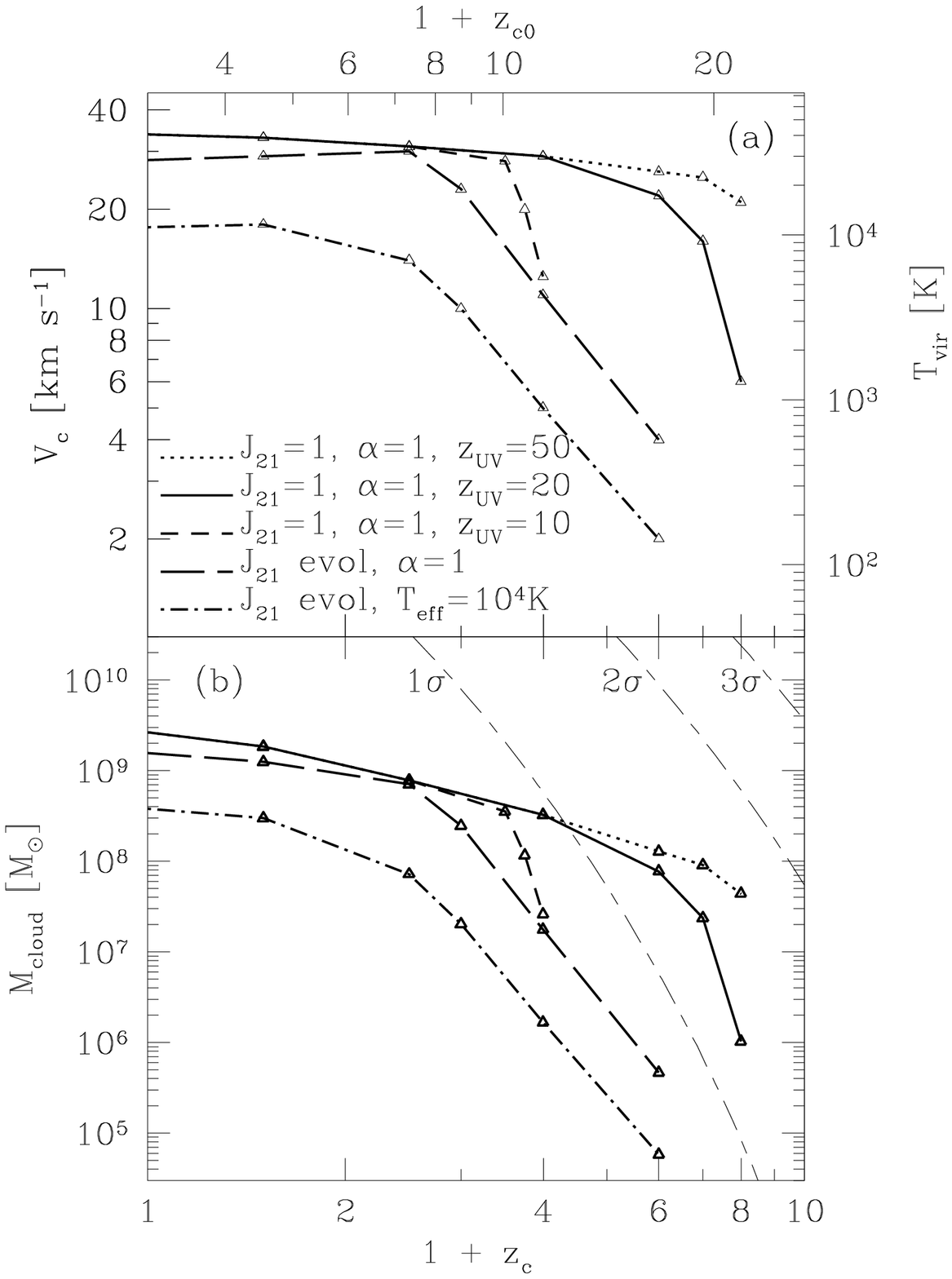,width=14cm}
\end{center}
\caption{(a) Critical circular velocity, and (b) critical baryon mass,
for H$_2$ cooling as a function of cloud collapse redshift $z_{\rm
c}$. Lines indicate models on the UVB evolution: $(J_{21},\alpha)=
(1,1)$ with various onset epochs, $z_{\rm UV}=50$ (dotted), 20 (solid)
and 10 (short dashed), and evolving $J_{21}$
(eq. [\protect\ref{eq:uvevl}\protect]) with different spectra,
$\alpha=1$ (long dashed) and $T_{\rm eff}=10^{4}$K (dot-dashed). Also
plotted in panel (b) are the baryon masses and collapse epochs of the
1,2,3$\sigma$ density perturbations in a $\Lambda$CDM model with
$(\Omega_0,\lambda_0,h_{100},\Omega_{\rm b},
\sigma_8)=(0.3,0.7,0.7,0.05,1.0)$.}
\label{fig:vz}
\end{figure*}
Alternatively, given a specific model on the evolution of the UVB, one
can deduce the critical circular velocity or mass for collapse and
cooling.  Fig. \ref{fig:vz} shows such quantities under which H$_2$
cooling is completely suppressed in an object with the cloud collapse
epoch $z_{\rm c}$. The UVB greatly prohibits cooling in clouds with
$V_{\rm c} \simlt 30$ km s$^{-1}$ or $M_{\rm cloud} \simlt 10^9 \msun$
at low redshifts. The threshold values, however, depend highly on the
time history and spectra of the UVB. For an evolutionary model given
by equation (\ref{eq:uvevl}), the critical mass drops by several
orders of magnitude at $z_{\rm c} \simgt 3$ ($z_{\rm c0} \simgt 10$),
enabling cooling in objects as small as $M_{\rm cloud} \sim 10^6
\msun$.

Fig. \ref{fig:vz}(b) also exhibits the mass and collapse redshift of
the 1,2,3$\sigma$ density perturbations in a cold dark matter (CDM)
model with $(\Omega_0,\lambda_0,h_{100},\Omega_{\rm b},
\sigma_8)=(0.3,0.7,0.7,0.05,1.0)$.  This is a specific example of the
most successful cosmological model which is consistent with a number
of current observations (e.g., Wang et al. 2000). Fig. \ref{fig:vz}
shows that the UVB mainly suppresses collapse and cooling of $\simlt 1
\sigma$ objects with $M_{\rm cloud} \simlt 10^9 \msun$ in the
$\Lambda$CDM model. It is rather difficult to suppress cooling in
higher-$\sigma$ objects, unless the radiation intensity becomes much
higher than the level observed at $z\sim 3$.

If different values of cosmological parameters are adopted, the
collapse criterion in terms of circular velocity remains practically
unchanged as long as $\Omega_{\rm b}/\Omega_0 \ll 1$. Thus, the
critical mass shown in Fig. \ref{fig:vz}(b) scales with cosmological
parameters merely according to equation (\ref{eq:vc}).

\section{Discussion}
\label{sec:dissc}

Our calculations predict that the formation of objects with $V_{\rm c}
\simlt 30$ km s$^{-1}$ is greatly suppressed by the UVB at low
redshifts, whereas much smaller objects ($T_{\rm vir} \simlt 10^4$K,
$V_{\rm c} \simlt 17$ km s$^{-1}$) can collapse and cool at high
redshifts ($z_{\rm c} \simgt 3$, $z_{\rm c0} \simgt 10$). This is
mainly because of three reasons: (i) intensity of the UVB tends to
decline at high redshifts prior to complete reionization of the
Universe; (ii) self-shielding of the gas is stronger at higher
redshifts owing to higher gas density; and (iii) the gas density in
Jeans unstable clouds rises rapidly along with run-away collapse,
further promoting self-shielding and H$_2$ formation.

What are the fates of low mass objects that are able to cool and
collapse in the presence of the UVB? If the cooled gas can promptly
turn into stars, they are likely to constitute `building blocks' of
present-day luminous structures.  While such objects may be detected
directly by future instruments such as {\it NGST}, they may also have
links to observed subgalactic objects (e.g., Pascarelle et al. 1996)
or dwarf galaxies. In contrast, those prevented to collapse by the UVB
will become gravitationally unbound and spread into intergalactic
medium.  They might contribute to Ly$\alpha$ absorption lines in the
QSO spectra (e.g., Umemura \& Ikeuchi 1984).

\begin{figure}
\begin{center}
   \leavevmode\psfig{figure=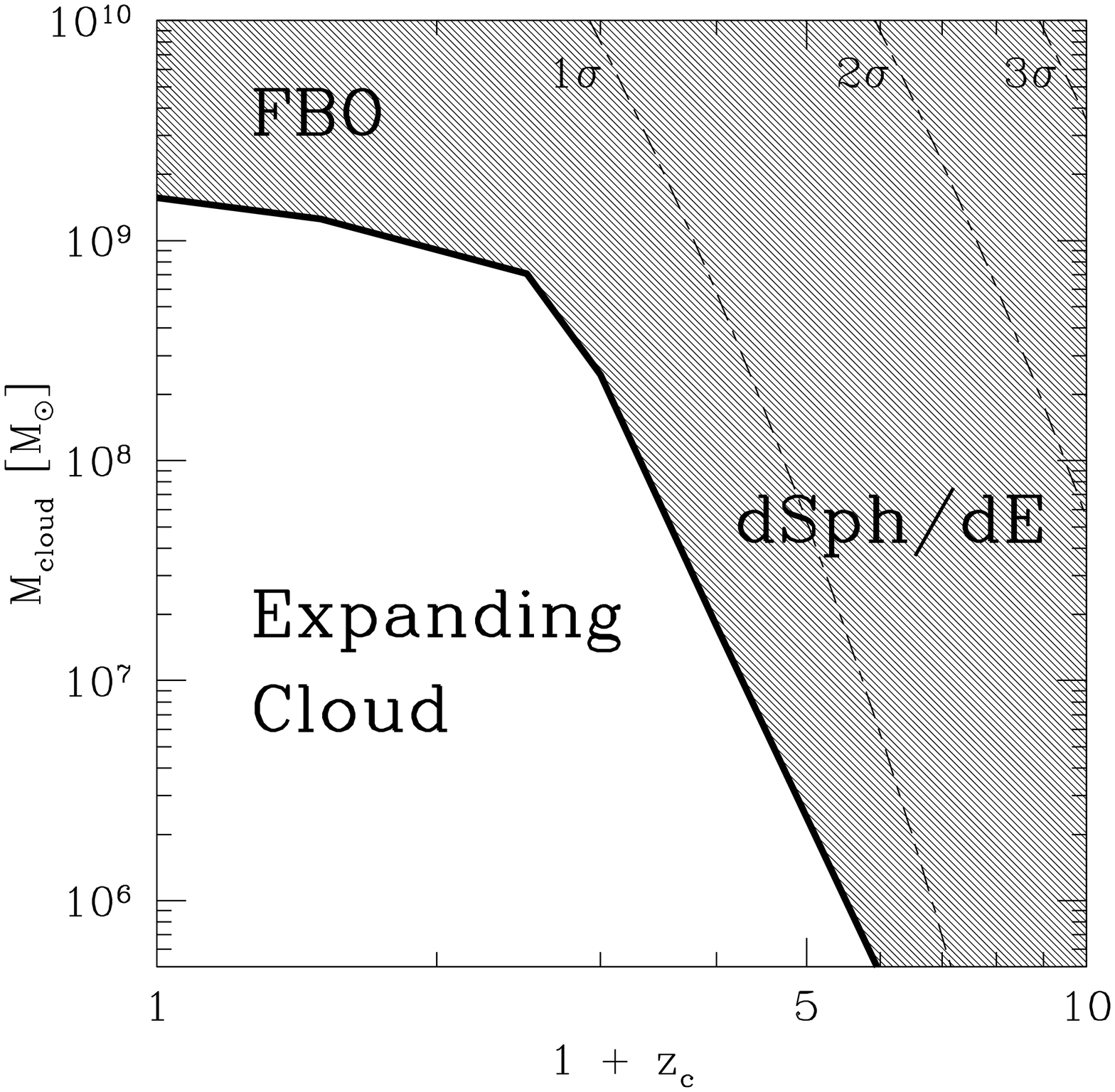,width=8.5cm}
\end{center}
\caption{Proposed bifurcation of low mass objects in the UVB. Objects
are classified into expanding clouds, dwarf spheroidal/elliptical
(dSph/dE) galaxies, and faint blue objects (FBOs), in terms of their
baryonic masses and collapse redshifts.  Thick solid line indicates
the critical mass for collapse and H$_2$ cooling in the case of
evolving $J_{21}$ (eq. [\protect\ref{eq:uvevl}\protect]) with
$\alpha=1$. Also plotted by thin lines are the masses and collapse
epochs of the 1,2,3$\sigma$ perturbations in the $\Lambda$CDM model.}
\label{fig:mbif}
\end{figure}
These speculations on the bifurcation of low mass objects are
illustrated schematically in Fig. \ref{fig:mbif}.  The critical
baryonic mass in the case of evolving $J_{21}$ (eq.~[\ref{eq:uvevl}])
with $\alpha=1$ is used to distinguish the uncollapsed `expanding
clouds' from the others. Furthermore, we expect that the collapsed
objects may be identified, mainly based on their distinct star
formation histories, as at least two separate populations: the
lower-mass ($\simlt 10^{9}\msun$), higher-$\sigma$ ($\simgt 1 \sigma$)
systems and the higher-mass, lower-$\sigma$ ones. The former
population can be a plausible candidate of dwarf spheroidal/elliptical
galaxies for the following reasons. First, star formation in these
objects is expected to be active at redshifts higher than $\sim 3$,
consistent with the fact that the majority of local group dwarfs show
a sharp decline in the star formation rate at some $10$ Gyr ago (Mateo
1998; Gnedin 2000a). Second, their gravitational potential is
compatible with small velocity dispersions ($\simlt 10$ km s$^{-1}$)
of local group dwarfs. Third, they are apt to be dark matter dominated
and metal-poor since the gas infall and star formation are suppressed
particularly at the outer envelope. Fourth, our simulations indicate
that the central gas in these objects can cool efficiently by H$_2$ in
the course of spherical contraction before reaching the rotation
barrier. Galaxy formation is thus likely to proceed rapidly in a
dissipationless manner and lead to the formation of spheroids rather
than disks.

On the other hand, the latter population will exhibit signs of more
recent star formation and contribute significantly to the global star
formation rate of the Universe. The clustering of these low $\sigma$
objects is expected to be rather weak. They might thus have been
identified as a part of faint blue objects (e.g., Tyson 1988; Cowie et
al. 1988; Ellis 1997) or star-forming dwarfs (e.g., Thuan \& Martin
1981; Lee et al. 2000). This is in line with the idea proposed by
Babul \& Rees (1992) and extended by Babul \& Ferguson (1996) and
Kepner et al. (1997). One distinction is that the impacts of the
decline of the UV intensity at low redshifts turn out to be
modest. This is because the collapse is delayed by the kinetic energy
of the gas attained during the peak of the UV intensity at $z\simeq 3 -
5$ (although the threshold circular velocity does decrease with time,
the threshold mass continues to rise due to a factor $M/V_{\rm c}^3
\sim [1+z_{\rm c}]^{-3/2}$).

The present spherical collapse simulations are mainly applicable to
low mass objects near the Jeans scale. Much larger objects, in
contrast, are likely to first undergo sheet-like pancake collapse and
evolve in a different manner from the spherical case. They would not
end up with the run-away collapse as gravity of the sheet is readily
overwhelmed by thermal pressure. In fact, Susa \& Umemura (2000a,b)
argued that the pancakes collapsing in the UVB can bifurcate,
depending upon the degree of self-shielding, into observed early and
late type galaxies. Both their results and ours suggest that the UVB
could play an essential role in regulating the star formation history
of the Universe.

Recent three-dimensional simulations show that a number of subclumps
tend to arise from the fragmentation process in a parent cloud (e.g.,
Bromm, Coppi \& Larson 1999; Abel, Bryan \& Norman 2000). The criteria
obtained in the current calculations are applicable to each of such
locally collapsing subclumps.  As a net effect to the parent cloud,
the gas clumpiness is likely to enhance self-shielding against the
external UVB, since the number of photons necessary to ionize the gas
increases roughly by a factor $\langle n_{\rm H}^2 \rangle/\langle
n_{\rm H} \rangle^2$ (Gnedin \& Ostriker 1997; Ciardi et al. 2000b).

Major uncertainties in our predictions are regarding the feedback from
the sources formed in an collapsing object.  While our simulations can
probe directly the cloud evolution until its central collapse, the
evolution thereafter will be influenced by several additional
processes. For instance, Omukai \& Nishi (1999) argued that even a
single OB star can photodissociate surrounding hydrogen molecules
within a static cloud with $T_{\rm vir} \simlt 10^4$K. On the other
hand, Ferrara (1998) pointed out that blow-away processes during
supernova explosions can enhance the fraction of hydrogen molecules in
surrounding media. Such explosions may also disrupt low mass objects
and induce metal injections into intergalactic space (e.g., Dekel \&
Silk 1986; Nishi \& Susa 1999; Ferrara \& Tolstoy 2000). It still
remains uncertain to what extent these processes affect successive gas
infall and cooling in a dynamically contracting object whose central
overdensity exceeds $\sim 10^5$.  As this issue is beyond the scope of
the present paper, we will investigate it separately in our future
publications.

\section{Conclusions}
\label{sec:concl}

We have studied the formation of Population III objects exposed to the
external UV radiation.  Radiative transfer of photons, H$_2$ formation
and destruction processes, and hydrodynamics are incorporated
explicitly in our analyses and all shown to play essential roles in
the development of Population III objects. Although the UVB does
suppress the formation of low mass objects, the negative feedback
turns out to be weaker than previously suggested in the following two
respects.

First, the cut-off scale of the collapse drops significantly below
$T_{\rm vir} = 10^4$K ($V_{\rm c} = 17$ km s$^{-1}$) at weak UV intensities
($J_{21}\simlt 10^{-2}$), due to both self-shielding of the gas and
H$_2$ cooling. Their importance becomes greater at higher redshifts.
At stronger intensities, objects as large as $V_{\rm c} \sim 40$ km s$^{-1}$
can be photoevaporated and prohibited to collapse, in agreement with
previous investigations based on the optically thin approximation
(e.g., Thoul \& Weinberg 1996). High mass objects well above this
scale can collapse almost irrespectively of the UVB by atomic cooling.

Secondly, a spherical cloud that has become Jeans unstable in the UVB
tends to undergo run-away collapse and does not settle into
hydrostatic equilibrium. As a result, even if the gas is ionized
and/or H$_2$ molecules are destroyed once by radiation at low
densities (e.g., Haiman et al. 2000), the radiation is attenuated and
H$_2$ can re-form during the dynamical collapse. The critical
densities at which self-shielding and efficient H$_2$ cooling take
place depend sensitively on the spectrum and intensity of the external
UVB. For a QSO-like ($\alpha=1$) spectrum, H$_2$ formation is even
enhanced by a little radiation with $J_{21}\sim 10^{-3}-10^{-2}$
(Haiman et al. 1996b). For a stellar-like ($T_{\rm eff}=10^4$K)
spectrum, H$_2$ formation is hindered to a greater extent by the
photons with energies below 13.6 eV (Haiman et al. 1997). In any case,
the collapsing gas is shown to cool efficiently to temperatures well
below $10^4$ K before rotationally supported, as long as the radiation
intensity is comparable to or less than $J_{21}\sim 1$ observed at
$z\sim 3$. The H$_2$ fraction in the cooled gas reaches the level
$\sim 10^{-3}$, comparable to that achieved commonly in the
metal-deficient post-shock layer (Shapiro \& Kang 1987; Ferrara 1998;
Susa et al. 1998).

Our results imply that star formation can take place in low mass
objects collapsing in the UVB. The threshold baryon mass for the
complete suppression of collapse and star formation is $M_{\rm cloud}
\sim 10^9 \msun$ at $z_{\rm c} \simlt 3$ but drops significantly at
higher redshifts. In a conventional $\Lambda$CDM universe, it
coincides roughly with that of the 1$\sigma$ density fluctuations at
$z_{\rm c} \simgt 3$. Objects near and above this threshold can thus
constitute building blocks of present-day luminous structures, and may
also have links to dwarf spheroidal/elliptical galaxies and faint blue
objects. These results indicate that the UVB can play a key role in
regulating the star formation history of the Universe.

\section*{Acknowledgments}
We thank Kuniaki Masai, Taishi Nakamoto and Yukiko Tajiri for
discussions, and the referee, Andrea Ferrara, for helpful comments. TK
and HS acknowledge support from Research Fellowships of the Japan
Society for the Promotion of Science for Young Scientists and Research
Grant of University of Tsukuba Research Project, respectively. This
work was partially carried out with facilities at the Center for
Computational Physics at University of Tsukuba.


\label{lastpage}
\end{document}